\documentclass[fleqn,usenatbib]{mnras}

\usepackage{newtxtext,newtxmath}

\usepackage[T1]{fontenc}
\usepackage{ae,aecompl}

\usepackage{graphicx}	
\usepackage{amsmath}	
\usepackage{amssymb}	
\usepackage[usenames,dvipsnames]{xcolor} 
\usepackage[normalem]{ulem} 
\usepackage{hyperref}

\usepackage[utf8]{inputenc}

\usepackage{lipsum} 
\usepackage{multirow}

\newcommand{\eagle}{\mbox{\sc{Eagle}}}

\newcommand{\flares}{\mbox{\sc Flares}}

\newcommand{\hb}{\mbox{H$\beta$}}

\newcommand{\oiii}{[\ion{O}{III}]}

\newcommand{\spitzer}{\mbox{\it Spitzer}}

\newcommand{\hubble}{\mbox{\it Hubble}}
\newcommand{\jwst}{\mbox{\it JWST}}

\title[FLARES XI: \ion{O}{III} emitters]{First Light And Reionisation Epoch Simulations (FLARES) XI: \newline[\ion{O}{III}] emitting galaxies at $5<z<10$}

\author[Stephen M. Wilkins et al.]{Stephen M. Wilkins$^{1,2}$\thanks{E-mail: s.wilkins@sussex.ac.uk}, 
Christopher C. Lovell$^{3}$, 
Aswin P. Vijayan$^{4,5}$, 
Dimitrios Irodotou$^{6}$,  
\newauthor
Nathan J. Adams$^{7}$, 
William J. Roper$^{1}$, 
Joseph Caruana$^{8,2}$, 
Jorryt Matthee$^{9}$, 
Louise T. C. Seeyave$^{1}$, 
\newauthor
Christopher J. Conselice$^{7}$,
Pablo G. P\'erez-Gonz\'alez$^{10}$,
Jack C. Turner$^{1}$, 
James M. S. Donnellan$^{1}$ 
\newauthor
Aprajita Verma$^{11}$, 
J. A. A. Trussler$^{7}$ 
\\
$^{1}$Astronomy Centre, University of Sussex, Falmer, Brighton BN1 9QH, UK\\
$^{2}$Institute of Space Sciences and Astronomy, University of Malta, Msida MSD 2080, Malta \\
$^{3}$Institute of Cosmology and Gravitation, University of Portsmouth, Burnaby Road, Portsmouth, PO1 3FX, UK\\
$^{4}$Cosmic Dawn Center (DAWN) \\
$^{5}$DTU-Space, Technical University of Denmark, Elektrovej 327, DK-2800 Kgs. Lyngby, Denmark \\
$^{6}$Department of Physics, University of Helsinki, Gustaf Hällströmin katu 2, FI-00014, Helsinki, Finland\\
$^{7}$Jodrell Bank Centre for Astrophysics, University of Manchester, Oxford Road, Manchester, UK\\
$^{8}$Department of Physics, Faculty of Science, University of Malta, Msida MSD 2080, Malta\\
$^{9}$ Department of Physics, ETH Z\"urich, Wolfgang-Pauli-Strasse 27, 8093 Z\"urich, Switzerland\\
$^{10}$ Centro de Astrobiolog\'{\i}a (CAB), CSIC-INTA, Ctra. de Ajalvir km 4, Torrej\'on de Ardoz, E-28850, Madrid, Spain\\
$^{11}$ Sub-department of Astrophysics, University of Oxford, Denys Wilkinson Building, Keble Road, Oxford, OX1 3RH, UK 
}

\date{Accepted XXX. Received YYY; in original form ZZZ}

\pubyear{2023}

\begin{document}
\label{firstpage}
\pagerange{\pageref{firstpage}--\pageref{lastpage}}
\maketitle

\begin{abstract}
\jwst\ has now made it possible to probe the rest-frame optical line emission of high-redshift galaxies extending to $z\approx 9$, and potentially beyond. To aid in the interpretation of these emerging constraints, in this work we explore predictions for [\ion{O}{III}]$\lambda\lambda4960,5008$\AA\ emission in high-redshift galaxies using the First Light and Reionisation Epoch Simulations (\flares). We produce predictions for the \oiii\ luminosity function, its correlation with the UV luminosity, and the distribution of equivalent widths (EWs). We also explore how the \oiii\ EW correlates with physical properties including specific star formation rate, metallicity, and dust attenuation. Our predictions are largely consistent with recent observational constraints on the luminosity function, average equivalent widths, and line ratios. However, they fail to reproduce the observed tail of high-EW sources and the number density of extreme line emitters. Possibilities to explain these discrepancies include an additional source of ionising photons and/or greater stochasticity in star formation in the model or photometric scatter and/or bias in the observations. With \jwst\ now rapidly building larger samples and a wider range of emission lines the answer to this remaining discrepancy should be available imminently. 
\end{abstract}




\section{Introduction}\label{sec:intro}

With the successful commissioning of \jwst, the detailed study of rest-frame optical emission of distant, high-redshift ($z>3$) galaxies has now opened up. Already, dozens of arcmin$^2$ of deep $>2$~\textmu m imaging have been obtained \citep[e.g.][]{Bagley22}. This has allowed us to identify samples of $z>10$ galaxies for the first time \citep[e.g.][]{Adams22,Atek22,Castellano22,Donnan22, Finkelstein22c, Naidu22a}, probe the rest-frame optical spectral energy distributions at $z>6$ \citep[e.g.][]{Adams22}, and study optical morphologies to high-redshift \citep[e.g.][]{Ferreira22a, Ferreira22b, Kartaltepe22}. At the same time, the first spectroscopic constraints have emerged making use of both \jwst's multi-object (MOS) and wide field slitless spectroscopic (WFSS) modes to study distant, high-redshift galaxies \citep[e.g.][]{Kashino22, Matthee23, Sun23, Tacchella22, Trump22, Trussler22, Curti23, Katz23}. With many more spectroscopic observations underway, large samples will soon emerge, providing new insights into the physical properties of distant galaxies as well as accurately constraining their redshifts.

The primary target of spectroscopic studies of distant galaxies is the various rest-frame optical emission lines. These not only permit an unambiguous determination of the redshift, but carry a wealth of information about the source of ionising photons and the composition and properties of the interstellar medium (ISM) in these distant galaxies.   

While a range of optical lines are now potentially accessible, in this work we focus on the \oiii$\lambda\lambda 4960,5008$\AA\ (hereafter \oiii) doublet. In this work we consider both the combined line luminosity and EW denoted by \oiii\ and individual lines denoted by e.g. \oiii$\lambda$5008\AA. The combined \oiii\ line flux is, at least for $Z\approx 0.001-0.01$, similar to H$\alpha$ while also being accessible at higher-redshift, $z\approx 9$ (c.f. $z\approx 6.6$ for H$\alpha$) with \jwst's near-infrared instruments alone. For typical metallicities this line is correlated with the ionising photon production, though is also sensitive to extreme metallicities ($Z<0.001$, $Z>0.01$) and the conditions of the ISM. Crucially, observational constraints have already emerged for statistically useful samples of galaxies \citep[e.g.][]{Sun23, Matthee23}.

To fully realise the constraining potential of these observations, in this work we make predictions for the \oiii\ properties of galaxies at $z=5-10$ using the First Light And Reionisation Epoch Simulations \citep[\flares][]{FLARES-I}. \flares\ is a suite of hydrodynamical simulations employing the \eagle\ physics model \citep{schaye_eagle_2015, crain_eagle_2015}, but with a strategy designed to efficiently extend the range of masses and luminosities simulated relative to the original \eagle\ reference simulation. \flares\ has also been used to study the evolution of galaxy sizes \citep{FLARES-IV, FLARES-IX}, colours \citep{FLARES-VI}, star formation and metal enrichment histories \citep{FLARES-VII}, the emergence of passive galaxies \citep{FLARES-VIII}, and galaxy bias \citep{FLARES-X}. This work builds on earlier efforts to model the nebular line emission in large samples of simulated high-redshift galaxies \citep[e.g][]{Wilkins2013d, Wilkins20, FLARES-II}. In this work we explore predictions for a range of properties including the \oiii\ luminosity function, the correlation of $L_{\oiii}$ with the UV luminosity, the equivalent width distribution, the correlation with physical properties, and the impact of dust attenuation.  We also compare these predictions with recent observational constraints.

This paper is structured as follows: in Section \ref{sec:theory} we explore the key physics driving \oiii\ emission in star forming galaxies, including the dependence on the star formation and metal enrichment history (\S\ref{sec:theory:sfzh}), geometry (\S\ref{sec:theory:geometry}), dust (\S\ref{sec:theory:dust}), and the assumed initial mass function and stellar population synthesis model (\S\ref{sec:theory:spsimf}). In Section \ref{sec:flares} we describe the \flares\ project, including details of our spectral energy distribution modelling procedure (\S\ref{sec:flares:sed}). Then in Section \ref{sec:predictions} we present our predictions for the \oiii\ properties of galaxies in \flares. In this section we also explore the impact of dust  (\S\ref{sec:predictions:dust}) and the correlation of \oiii\ emission with key physical properties  (\S\ref{sec:predictions:physical}). In Section \ref{sec:observations} we then compare our predictions with recent observations from \hubble\ and \jwst. We then summarise our findings in this work and present our conclusion in Section \S\ref{sec:conc}. 


\section{Theoretical Background}\label{sec:theory}

Ionising photons produced by massive stars, active galactic nuclei (AGN), or other phenomena (e.g. shocks) lead to the formation of \ion{H}{II} regions. Within these regions various physical processes result in the formation of nebular line and continuum emission. The strength of individual emission lines are sensitive to the shape and normalisation of the ionising spectrum, alongside the properties of the ionised region, including its geometry, composition, and other physical properties. 

Amongst the most prominent and useful lines is the \oiii\ doublet. In this section we explore how the luminosities and equivalent widths of the [\ion{O}{III}] doublet are affected by the star formation history, metallicity, and geometry. To do this we employ smooth parametric star formation histories, a single metallicity shared by both the stellar population and surrounding gas, and a simple screen model for reprocessing by dust and gas (i.e. stellar populations are all equally affected by dust). Initially, we assume the following fiducial parameters and model choices: to model the stellar emission we use the v2.2.1 of the Binary Population And Spectral Synthesis \citep[BPASS][]{BPASS2.2.1} stellar population synthesis (SPS) model and  a \citet{chabrier_galactic_2003} initial mass function (IMF), while to model the nebular emission we use version 17.03 of the \texttt{cloudy} photo-ionisation code \citep{Ferland2017} and assume a reference ionising parameter ($U_{\rm ref}$) of $0.01$, a solar abundance pattern, and no escape of ionising photons. Most of these assumptions are explored in this section.   

\subsection{Star formation and metal enrichment history}\label{sec:theory:sfzh}

\begin{figure}
	\includegraphics[width=\columnwidth]{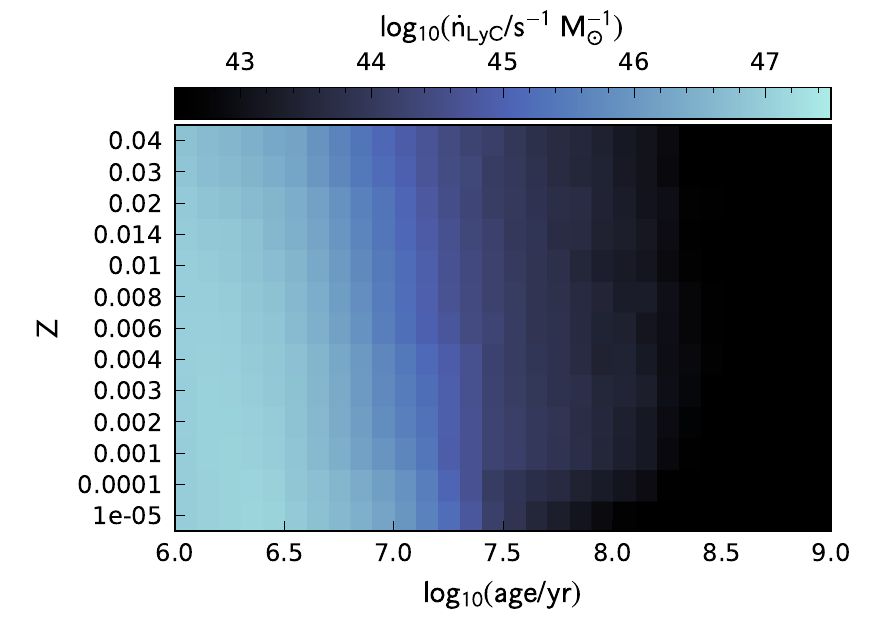}
	\caption{The specific ionising photon luminosity of a simple stellar population as a function of age and metallicity assuming BPASS v2.2.1 and \citet{chabrier_galactic_2003} IMF with $m_{\rm up}=300\ {\rm M_{\odot}}$.\label{fig:theory_log10Q}}
\end{figure}

In a stellar population the production of Lyman-continuum (LyC) photons is dominated by hot, massive, and short lived stars. Consequently, the LyC luminosity ($\dot{n}_{\rm LyC}$) drops precipitously as the stellar population ages. This is demonstrated in Figure \ref{fig:theory_log10Q}, where we show the \emph{specific} LyC luminosity, i.e. the LyC luminosity per unit stellar mass, as a function of age and metallicity. This reveals that the LyC luminosity drops by a factor of $\approx 10^{4}$ as a population ages from $t=1\to 100\, {\rm Myr}$. As lower-metallicity stars can attain higher temperatures, the LyC luminosity is higher at low metallicity, at least at young ages ($<10\ {\rm Myr}$). For older stellar populations this trend reverses as the most massive stars evolve off the main sequence faster.

\begin{figure}
	\includegraphics[width=\columnwidth]{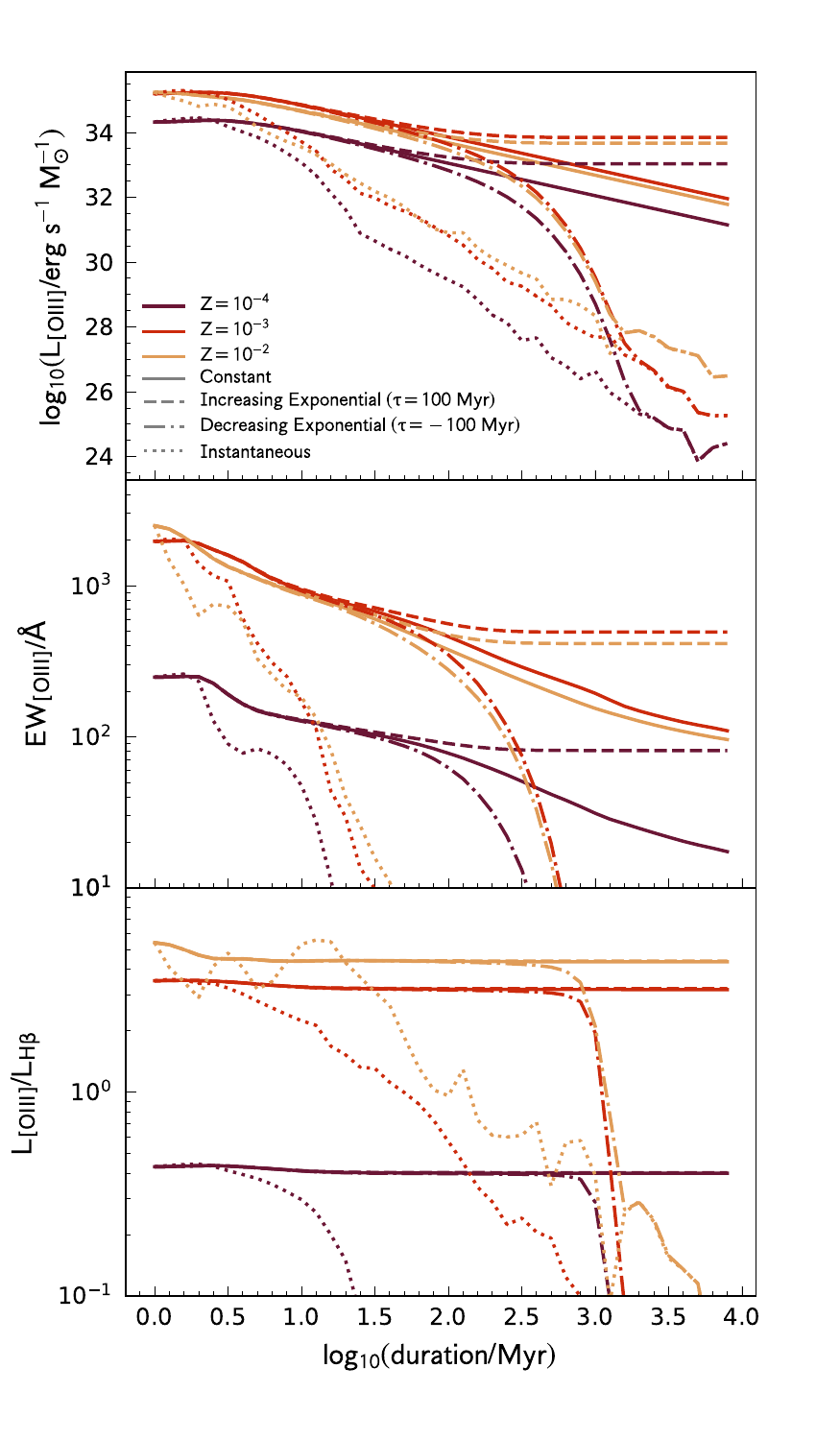}
	\caption{The specific \oiii\ luminosity, \oiii\ equivalent width, and of \oiii/\hb\ luminosity ratio as a function of the duration of star formation assuming four different star formation histories: an instantaneous burst (dotted line), exponentially decreasing star formation (dot-dashed line), constant star formation (solid line), and exponentially increasing star formation (dashed line), and three metallicities: $Z=10^{-4},\ 10^{-3}, 10^{-2}$. \label{fig:theory_sfh}}
\end{figure}

The consequence of this strong dependency on age and metallicity is that any emission line is strongly sensitive to the star formation and metal enrichment history. In Figure \ref{fig:theory_sfh} we explore how the \emph{specific} (i.e. per unit stellar mass formed) \oiii\ luminosity, equivalent width (EW), and ratio to the \hb\ luminosity are affected by the star formation history. Here we present results assuming four simple star formation histories: an instantaneous burst, an exponentially declining ($\tau=-100$ Myr), constant, and exponentially increasing star formation ($\tau=100$ Myr). Unsurprisingly in each case the specific luminosity and EW drops, at least initially. With no replenishment of massive stars in the instantaneous model the luminosity and EW drop by four orders of magnitude over $t=1\to 100\, {\rm Myr}$. For the other models both drop more slowly, with the luminosity (EW) dropping by $\approx 10\times$ ($\approx 4\times$) over $t=1\to 100\, {\rm Myr}$. At this point the exponentially increasing model plateaus as the rapidly increasing star formation rate balances the accumulation of longer lived lower mass stars. Conversely in the exponentially declining model the luminosity and EW rapidly drop, eventually joining an instantaneous burst. In the constant model the specific luminosity and EW drop at a near constant fractional rate. For scenarios with continuing star formation the \oiii/\hb\ luminosity ratio is largely constant since the shape of the ionising spectrum remains roughly constant. Conversely, for an instantaneous burst the \oiii/\hb\ ratio drops as the population ages due to the changing shape of the ionising spectrum.

\begin{figure}
	\includegraphics[width=\columnwidth]{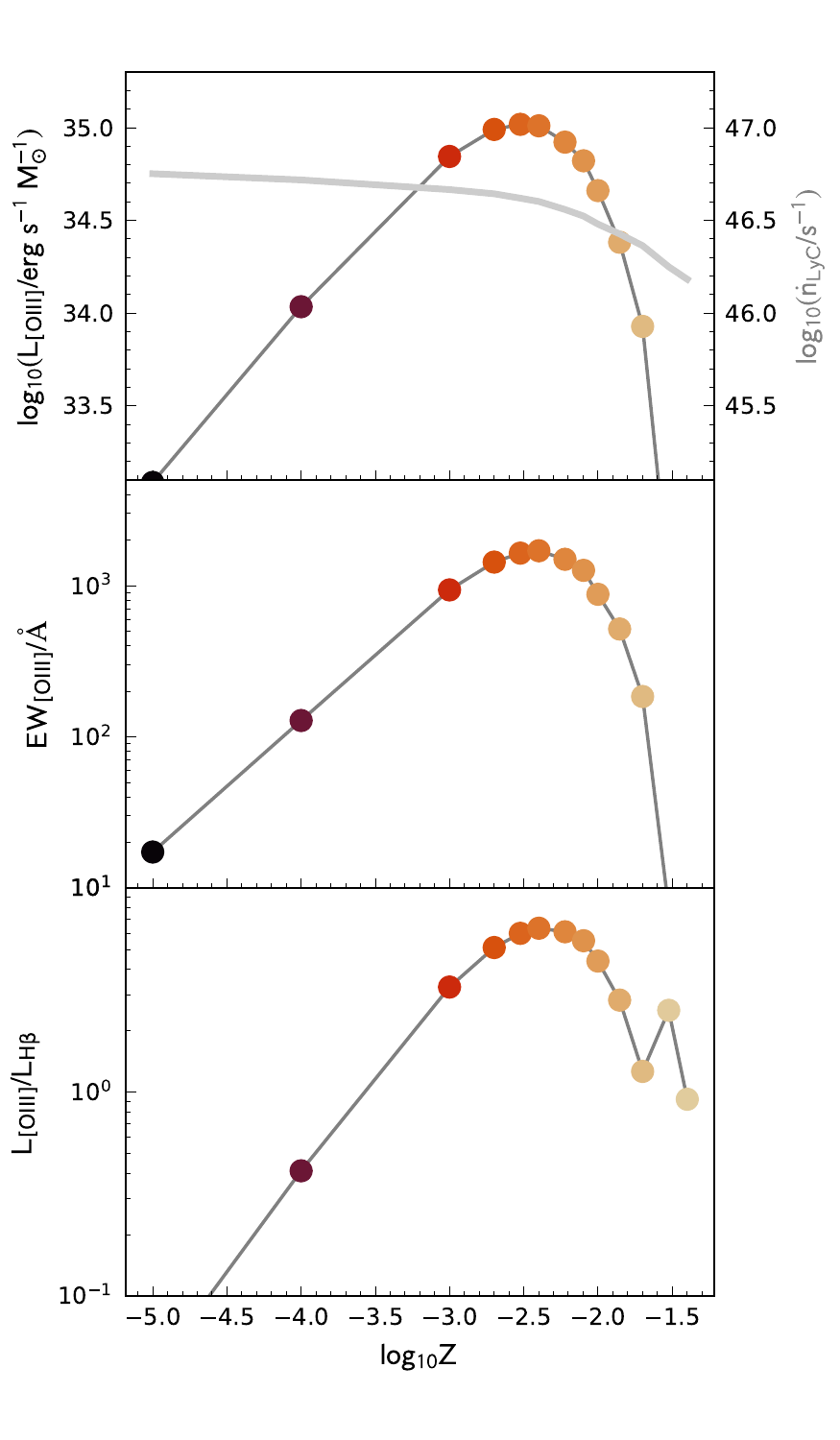}
	\caption{The specific \oiii\ luminosity, \oiii\ equivalent width, and of \oiii/\hb\ luminosity ratio as a function of metallicity assuming 10 Myr constant star formation. Also shown on the top panel is the sensitivity of the LyC luminosity.\label{fig:theory_Z}}
\end{figure}

Figure \ref{fig:theory_sfh} also presents predictions for each star formation history scenario for three different metallicities: $Z=10^{-4},\ 10^{-3},$ and $10^{-2}$. Metallicity has an impact through both changing the ionising continuum normalisation and shape, and changing the composition of the nebular region itself. To show this more clearly, in Figure \ref{fig:theory_Z} we show the luminosity, EW, and \oiii/\hb\ ratio as a function of metallicity assuming 10 Myr constant star formation. In the top (luminosity) panel we also show the LyC luminosity. While the LyC luminosity increases to lower metallicity, the \oiii\ luminosity (and thus EW) reaches a peak at $Z\sim 0.005$, dropping rapidly on either side. At low-metallicity the decrease is driven by dropping abundance of Oxygen, while the drop to high-metallicities is due to both the falling LyC luminosity and changing shape of the ionising spectrum.

\subsubsection{Composition}\label{sec:theory:composition}

\begin{figure}
	\includegraphics[width=\columnwidth]{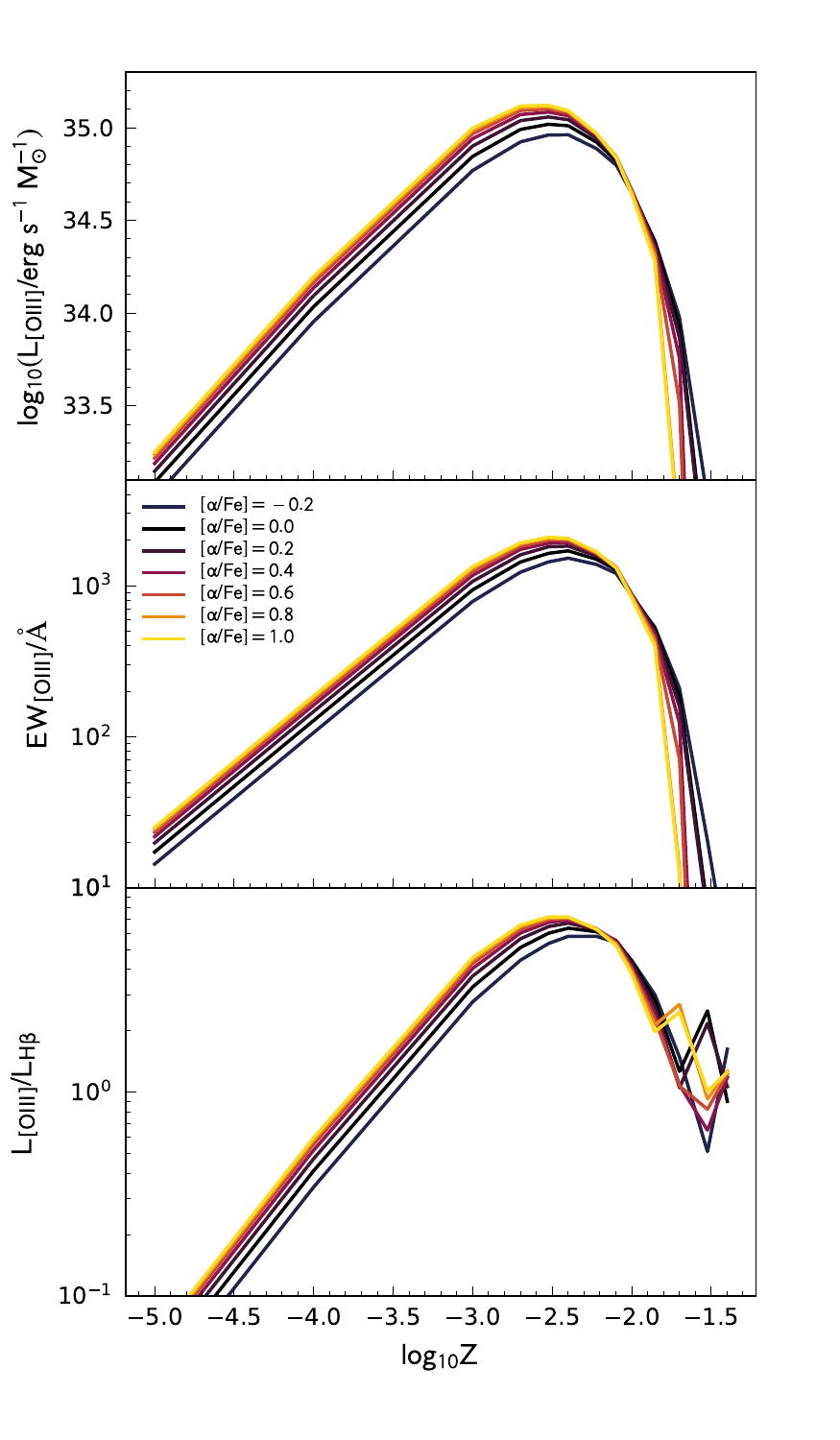}
	\caption{The same as Figure \ref{fig:theory_Z} but showing different levels of $\alpha$-enhancement in the nebular gas. \label{fig:theory_alpha}}
\end{figure}

In our modelling so far we have assumed that the composition of the nebular gas follows a scaled Solar abundance pattern. However, at high-redshift, galaxies will be increasingly enhanced with $\alpha$-elements due to the shorter timescale for their production \citep{Steidel2016}. To explore the impact of $\alpha$-enhancement in Figure \ref{fig:theory_alpha} we show the specific \oiii\ luminosity, \oiii\ equivalent width, and \oiii/\hb\ luminosity ratio as a function of metallicity for different $\alpha$-enhancements assuming 10 Myr constant star formation. Since the underlying stellar models assume Solar composition, this modelling is not self-consistent; nevertheless, it allows us to explore the potential impact. This reveals that boosting the enhancement to [$\alpha$/Fe]$=0.6$ boosts \oiii\ luminosities and EWs by $\approx 0.1$ dex at $Z<0.003$. At $Z> 0.005$ the effect becomes small and leads to a suppression at super-solar metallicities. 

\subsubsection{Geometry}\label{sec:theory:geometry}

In our \texttt{cloudy} modelling the geometry of the H\textsc{ii} region is encapsulated by the LyC escape fraction, the geometry, and the average ionisation parameter ($U$). $U$ encodes the the number of LyC photons per atom.

In our toy model the impact of the escape fraction is to simply scale line luminosities while \emph{almost} leaving the continuum unchanged. The continuum changes slightly due to the contribution of nebular emission to the continuum (which itself will scale with the escape fraction). 

In this work we follow earlier modelling and assume a spherical geometry, i.e. the ionisation parameter varies as $r^{-2}$ through the cloud. In this case the assumed ionisation parameter corresponds to the volume averaged (over the Str{\"o}mgren sphere) value ($\langle U\rangle$) which is approximately equivalent to $3U(r=R_s)$, where $R_s$ is the radius of the corresponding Str{\"o}mgren sphere. Since in a plane parallel geometry $U$ does not vary, the resulting line ratios can be different.

In the modelling thus far we have assumed a fixed reference ionisation parameter of $U_{\rm ref}=0.01$ referenced at $t=1$ Myr and $Z=0.01$. What this means is that the actual ionisation parameter assumed by our \texttt{cloudy} implementation varies according to the metallicity and age of the stellar population. In a spherical geometry the size of the Str{\"o}mgren sphere scales with cube-root of the ionising photon luminosity $Q$ thus the relationship between $U$, $Q$, and their reference values is:
\begin{equation}
    \langle U\rangle = \langle U\rangle_{\rm ref}\left(\frac{Q}{Q_{\rm ref}}\right)^{1/3}.
\end{equation}

\begin{figure}
	\includegraphics[width=\columnwidth]{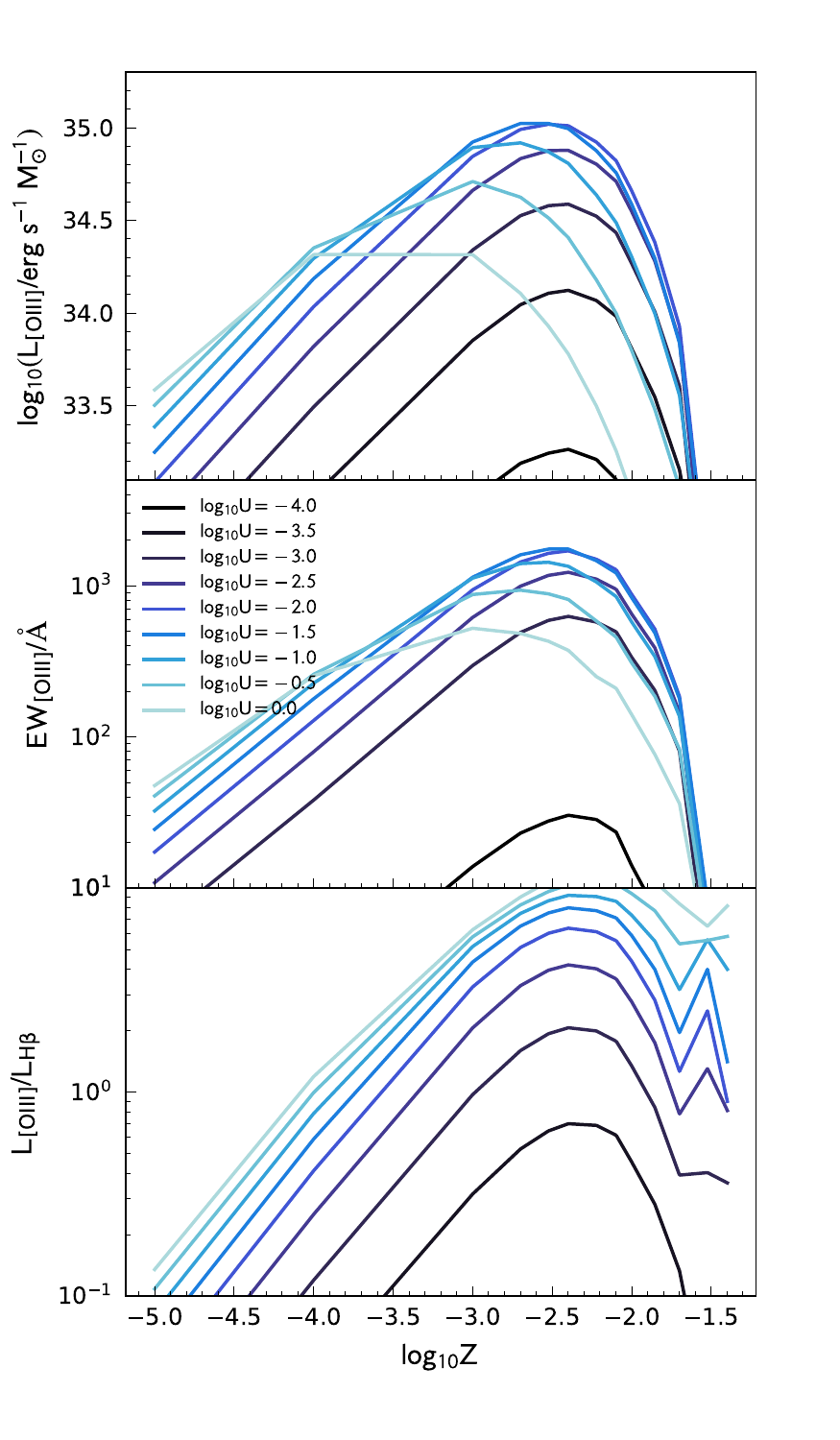}
	\caption{As Figure \ref{fig:theory_Z} but showing the impact of varying the reference ionisation parameter $U_{\rm ref}$.\label{fig:theory_U}}
\end{figure}

To explore how the ionisation parameter affects our predictions, in Figure \ref{fig:theory_U} we show the specific luminosity, EW, and \oiii/\hb\ line ratio as a function of metallicity but assuming several different reference ionisation parameters, $\log_{10}U_{\rm ref} = -4\to 0$. This reveals the complex sensitivity of \oiii\ to $U$. The \oiii\ mass-weighted luminosity peaks assuming $\log_{10}U_{\rm ref} \approx -2$ (i.e. our default value) at $Z=0.003$. The location of the peak is sensitive to the choice of $U$ shifting to lower metallicity with higher $U$. This shift is sufficient such that at very low metallicity ($Z<0.0001$) the \oiii\ emission is highest for higher values of $U$.

\subsubsection{Dust}\label{sec:theory:dust}

Like any other optical photons, the \oiii\ line emission is susceptible to dust attenuation. For a simple screen dust geometry, EWs will be unaffected by dust since both the continuum and line emission would be attenuated by the same amount. In reality, however, dust and stars have a more complex geometry, likely leaving EWs sensitive to dust attenuation. For example, a model in which young stars and their associated \ion{H}{II} regions are attenuated more strongly than older stars would naturally result in line emission suffering higher attenuation than the underlying continuum, reducing the equivalent width. On the other hand, if star formation is preferentially taking place on the outskirts of galaxies,  where the attenuation is lower, then the attenuation of the overall continuum could be higher, boosting EWs relative to their intrinsic values. This is an important consideration, since \flares\ includes both a birth cloud component and a wider attenuation for each star particle determined by the distribution of metals along the line-of-sight. In the context of the \flares\ predictions this is explored in \ref{sec:predictions:dust}.

\subsection{Initial Mass Function and Stellar Population Synthesis Model}\label{sec:theory:spsimf}

\begin{figure}
	\includegraphics[width=\columnwidth]{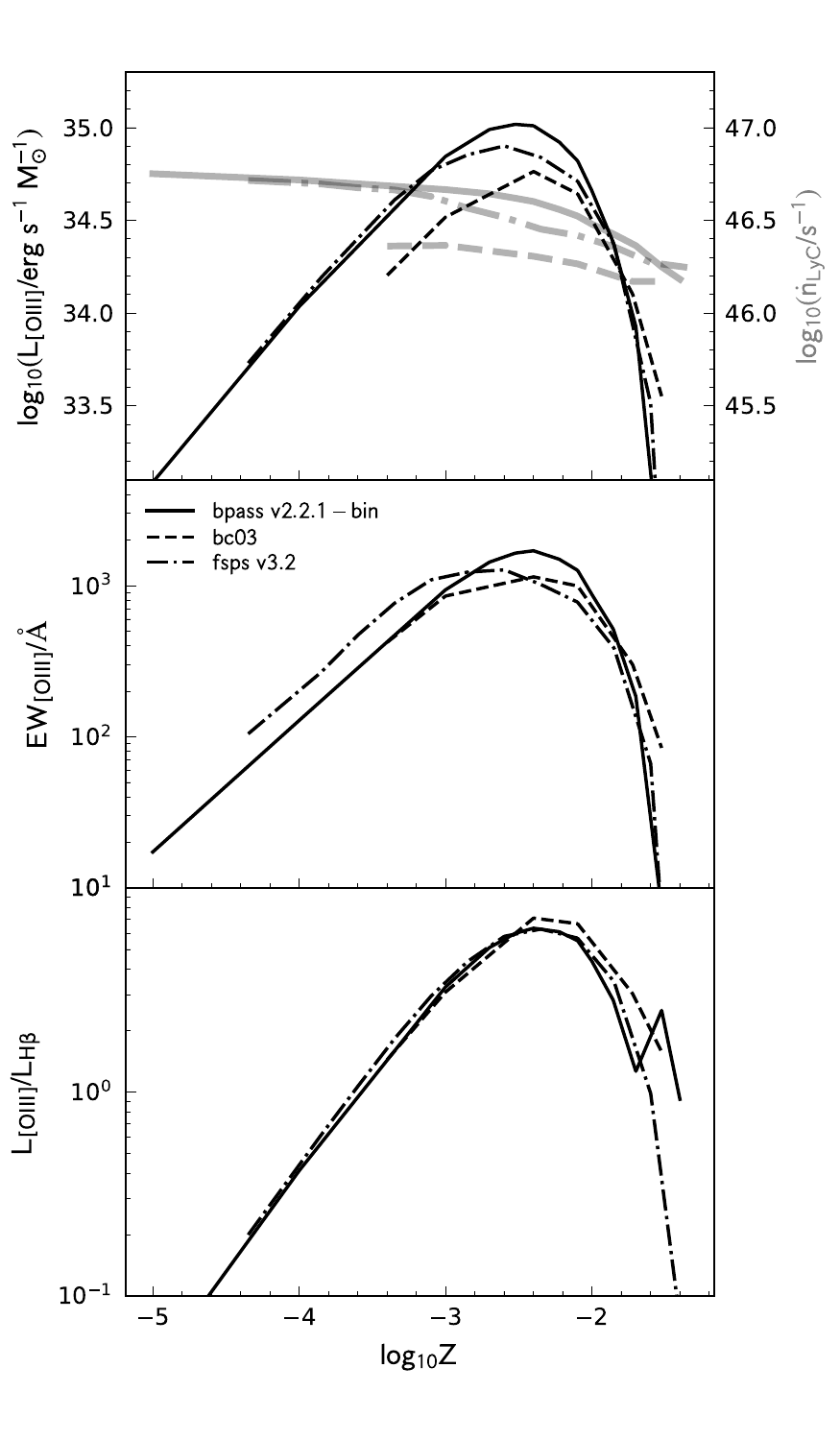}
	\caption{As Figure \ref{fig:theory_Z} but for two additional population synthesis models: FSPS and BC03. \label{fig:theory_sps}}
\end{figure}

Finally, any predicted observational quantity is also going to be affected by the choice of Initial Mass Function (IMF) and Stellar Population Synthesis (SPS) model. In Figure \ref{fig:theory_sps} we show the predicted luminosity and EW as a function of metallicity for three different SPS models, including BPASS (our default),  the Flexible Stellar Population Synthesis \citep[FSPS][]{FSPSI} code, and the \citet{BC03} (BC03) models. Due the inclusion of binary interactions the BPASS model generally yields higher LyC luminosities, resulting in higher line luminosities and EWs \citep{Stanway2016}. At $Z<0.001$ FSPS yields slightly higher luminosities, but EWs which are $\approx 50\%$ higher than BPASS due to fainter continuum emission.

\begin{figure}
	\includegraphics[width=\columnwidth]{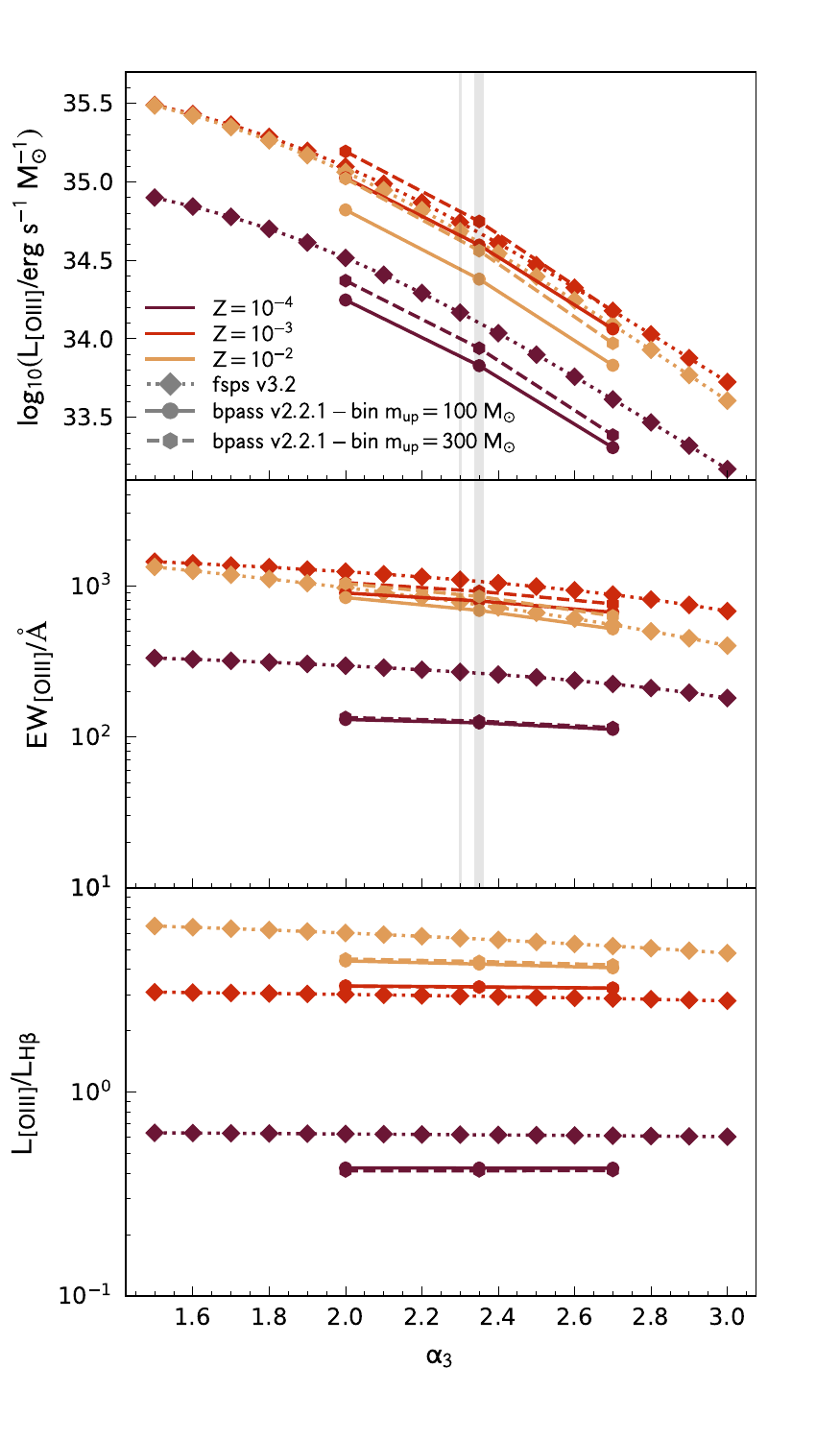}
	\caption{The dependence of the line luminosity and EW on the high-mass slope of the IMF $\alpha_{3}$ assuming BPASS and FSPS. The two vertical lines denote the high-mass slope assumed by the \citet{chabrier_galactic_2003} ($\alpha=2.3$) and \citet{Salpeter1955} ($\alpha=2.35$).}\label{fig:theory_imf}
\end{figure}

Since the IMF controls the relative proportions of stars and line emission is driven by the most massive stars the luminosity, and potentially EWs, will be sensitive to the shape of the IMF. In Figure \ref{fig:theory_imf} we show the predicted specific line luminosities and EWs as a function of the high-mass slope of the IMF ($\alpha_{3}$) using the BPASS and FSPS models which include this flexibility. In both FSPS and BPASS the parameter $\alpha_{3}$ describes the slope of the IMF at $>1\ {\rm M_{\odot}}$ (c.f. Salpeter: $\alpha=2.3$). We do this for three metallicities but in each case assume 10 Myr constant star formation. Unsurprisingly, this reveals that line luminosities are strongly impacted by the choice of $\alpha_{3}$ with the line luminosity increasing by $\approx 0.5$ dex for changes to the slope of 0.35. However, EWs are only subtly affected since the optical continuum emission is also enhanced.  

\section{First Light And Reionisation Epoch Simulations}\label{sec:flares}

In this study we make use of the First Light And Reionisation Epoch Simulations (\flares). \flares\ is introduced in \citet{FLARES-I} and \citet{FLARES-II} and we refer the reader to those papers and references therein for a detailed introduction. In brief, \flares\ is a suite of hydrodynamical re-simulations. The core\footnote{In addition to the core runs \flares\ includes a range of simulations exploring changes to the physics model.} \flares\ suite adopts the AGNdT9 variant of the \eagle\ simulation project \cite[][]{schaye2015_eagle,crain2015_eagle} with identical resolution to the \eagle\ reference run. The core suite consists of re-simulations (zoom simulations) of 40 regions selected from a large low-resolution $(3.2\ {\rm Gpc})^3$ dark matter only simulation. Each re-simulation is $14/h\ {\rm cMpc}$ in radius. The selected regions span a large range in over-density (at $z\approx 4.7$): $\delta + 1 \approx -1\to 1$, with over-representation of the extremes of the density distribution. 

The \flares\ strategy yields galaxies across a large range of galaxy and halo masses and across a wide range of environments. The resolution of \flares\ enables us to confidently simulate galaxies down to $M_{\star}\approx 10^8\ {\rm M_{\odot}}$ where each galaxy is resolved with at least 100 star particles. At $z=10$ the most massive galaxy has $M_{\star}\approx 10^{10.3}\ {\rm M_{\odot}}$ while at $z=5$ the most massive galaxy has $M_{\star}\approx 10^{11.3}\ {\rm M_{\odot}}$. These are approximately 10$\times$ larger than the most massive galaxies found in the \eagle\ reference simulation.

Because \flares\ combines 40 individual re-simulations of biased regions the galaxies in each individual region have to be appropriately weighted before constructing scaling relations and distribution functions. This is described in detail in \citet{FLARES-I}.

\subsection{Spectral Energy Distribution modelling}\label{sec:flares:sed}

The spectral energy distribution (SED) modelling of galaxies in \flares\ is described in depth in \citet{FLARES-II}. In short, we associate every star particle\footnote{The initial star particle mass is $\approx 2\times 10^{6}\ {\rm M_{\odot}}$.} in the simulation with a pure stellar SED based on its mass, age, and metallicity using v2.2.1 of the Binary Population And Spectral Synthesis \cite[BPASS]{BPASS2.2.1} stellar population synthesis (SPS) library, and assume a \cite[]{chabrier_galactic_2003} initial mass function (IMF).

\begin{figure*}
    \includegraphics[width=2\columnwidth]{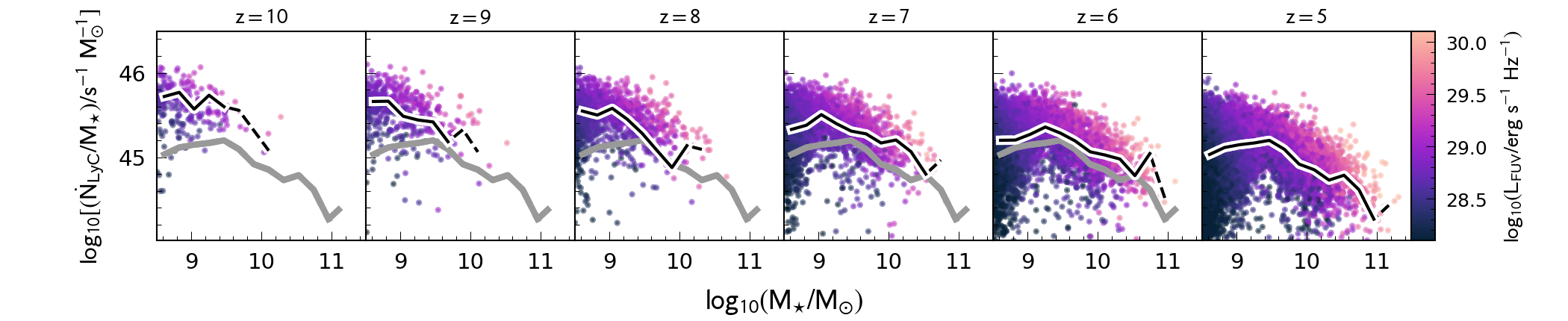}
    \includegraphics[width=2\columnwidth]{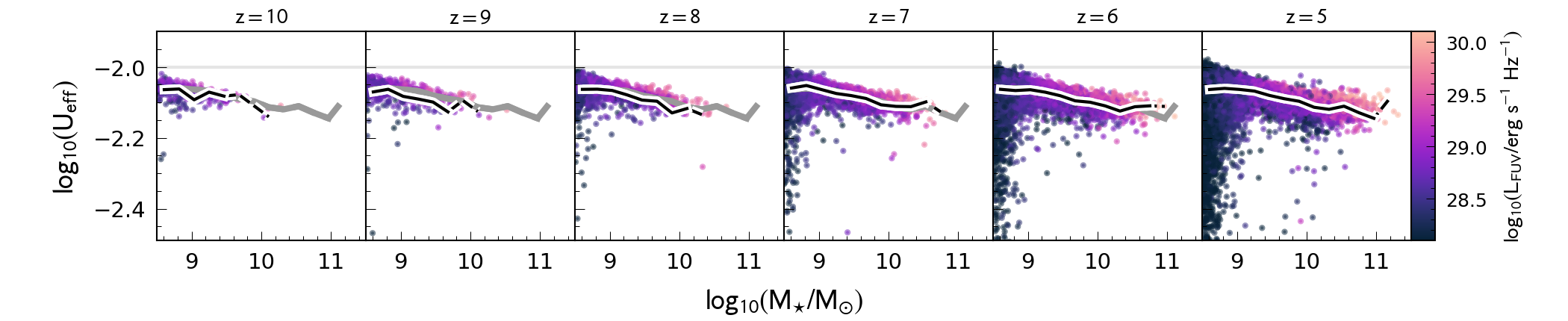}
	\caption{The evolution of the specific LyC photon production rate ($\dot{N}_{\rm LyC}/{\rm s^{-1}\ M_{\odot}^{-1}}$) [top] and effective ionisation parameter $U_{\rm eff}$ [bottom] from $z=5\to 10$ predicted by \flares. The effective ionisation parameter in this context is the combination of all star particles weighted by their LyC luminosity. The thick grey line denotes the $z=5$ relation in both figures. \label{fig:flares}}
\end{figure*}

In the top panel of Figure \ref{fig:flares} we present the \emph{specific} (i.e. per unit stellar mass) LyC photon production rate predicted for galaxies in \flares. This quantity simply depends on the star formation and metal enrichment histories of galaxies, since it is independent of reprocessing by dust and gas. This reveals a clear downward trend at high-masses ($M_{\star}>10^{9.5}\ {\rm M_{\odot}}$) but remains flatter at lower mass. This is primarily due to the strong evolution of stellar metallicities across this mass-range \citep[see][]{FLARES-VII} combined with the strong dependence of LyC photon production on metallicity (see \S\ref{sec:theory:sfzh}). Figure \ref{fig:flares} also reveals significant redshift evolution, with the specific production rate declining by $\approx 0.6$ dex from $z=10\to 5$. Since the mass-metallicity relationship evolves weakly with redshift \citep[see][]{FLARES-VII} over this range, this decrease reflects the changing star formation histories of galaxies, in particular the increase in average ages.

\subsubsection{Nebular emission modelling}

Once we have assigned a stellar SED we then associate each star particle with an ionisation bounded \ion{H}{II} region using version 17.03 of the \textsc{cloudy} photo-ionisation code \citep{Ferland2017}. Specifically, we use the pure stellar spectrum as the incident radiation field, assume the metallicity of the nebula is identical to the star particle, a solar abundance pattern, a covering fraction of 1 (corresponding to a LyC escape fraction of $\approx 0$ for an ionisation bound nebula) and a metallicity and age dependent ionisation parameter referenced at $t=1$ Myr and $Z=0.01$ of $U=0.01$. At other ages and metallicities the assumed ionisation parameter is scaled from this reference value using the ratio of the ionising luminosities of the two populations. A consequence of this is that nebular SEDs are formed from a wide range of ionisation parameters, with the maximum approximately that of the reference value. The bottom panel of Figure \ref{fig:flares} shows the effective (ionising photon luminosity weighted) ionisation parameter of galaxies in \flares. Unsurprisingly, these are clustered around our reference ionisation parameter, though show some modest decrease to higher masses, reflecting the shift to higher metallicities. The assumption of a solar abundance pattern was a decision made to align with the BPASS SPS library. However, galaxies at high-redshift are observed \citep[e.g][]{Cullen21} and predicted \citep[e.g][]{FLARES-VII} to be strongly enhanced with $\alpha$-elements, including Oxygen. \citet{FLARES-VII} found typical enhancements of [$\alpha$/Fe]$=0.6-0.8$ at $5<z<10$. Our modelling in \S\ref{sec:theory:composition} revealed that this level of $\alpha$-enhancement can boost the \oiii\ luminosity and EW fluxes by $\approx 0.1$ dex. In a future iteration of \flares\ we plan to address this self-consistently by using the newest version (v2.3) of the BPASS models which include $\alpha$-enhancement in the stellar atmosphere modelling. 

\subsubsection{Dust attenuation}

As described in \citet{FLARES-II} in \flares\ we implement a two component dust attenuation model. First, we associate young stellar populations (with age less than 10 Myr, following \citet{CF00} that birth clouds disperse along these timescales) with a metallicity dependent dusty birth cloud. Secondly, for each star particle (and associated \ion{H}{II} region) we apply attenuation due to dust in the intervening inter-stellar medium. This is determined by calculating the line-of-sight surface density of metals along the spatial $z$-axis, for each star particle, and converting this to an optical depth. For both components we assume a simple $\lambda^{-1}$ dependence of the attenuation. In \citet{FLARES-XII} we explore some of the wider features of this dust model. The impact of dust modelling is explored in \S\ref{sec:predictions:dust}.

\section{Predictions}\label{sec:predictions}

We now explore predictions for the \oiii\ properties of galaxies in \flares, including the \oiii\ luminosity function, correlation with UV luminosity, and equivalent width distribution. We then explore how our predictions are impacted by dust attenuation and how the \oiii\ EW correlates with other physical properties. 

\subsection{\oiii\ luminosity function}

\begin{figure*}
    \includegraphics[width=2\columnwidth]{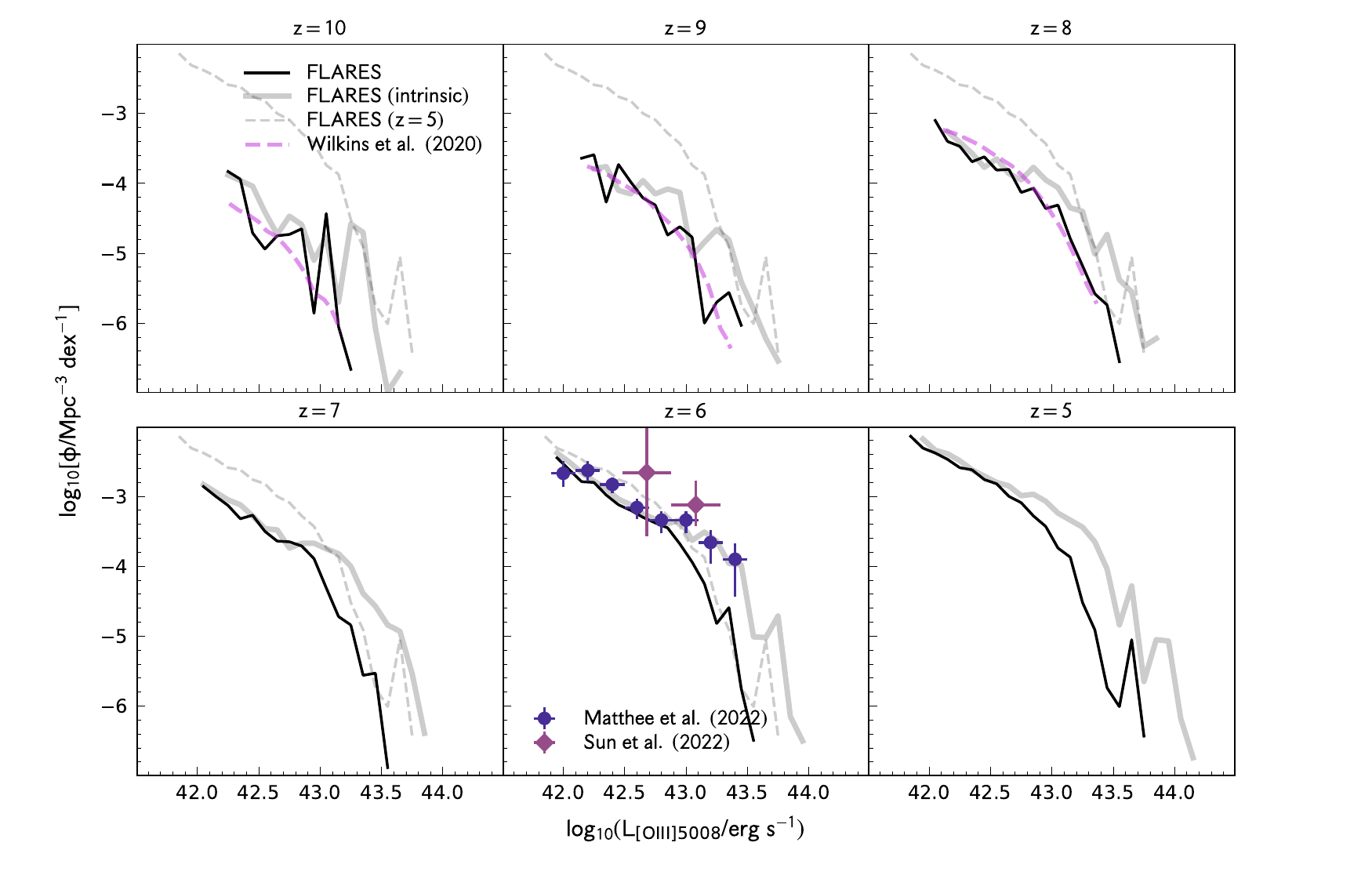}
	\caption{The evolution of the \oiii$\lambda 5008$\AA\ luminosity function from $z=5\to 10$ predicted by \flares. The dark thin line shows the \emph{observed} (dust-attenuated) luminosity function while the thicker fainter line shows the \emph{intrinsic} LF. The dashed line is the $z=5$ LF to highlight the evolution of the luminosity function. Observational constraints on the LF from \citet{Sun23} and \citet{Matthee23} are also shown at $z\approx 6$. \label{fig:LF}}
\end{figure*}

\begin{figure*}
    \includegraphics[width=2\columnwidth]{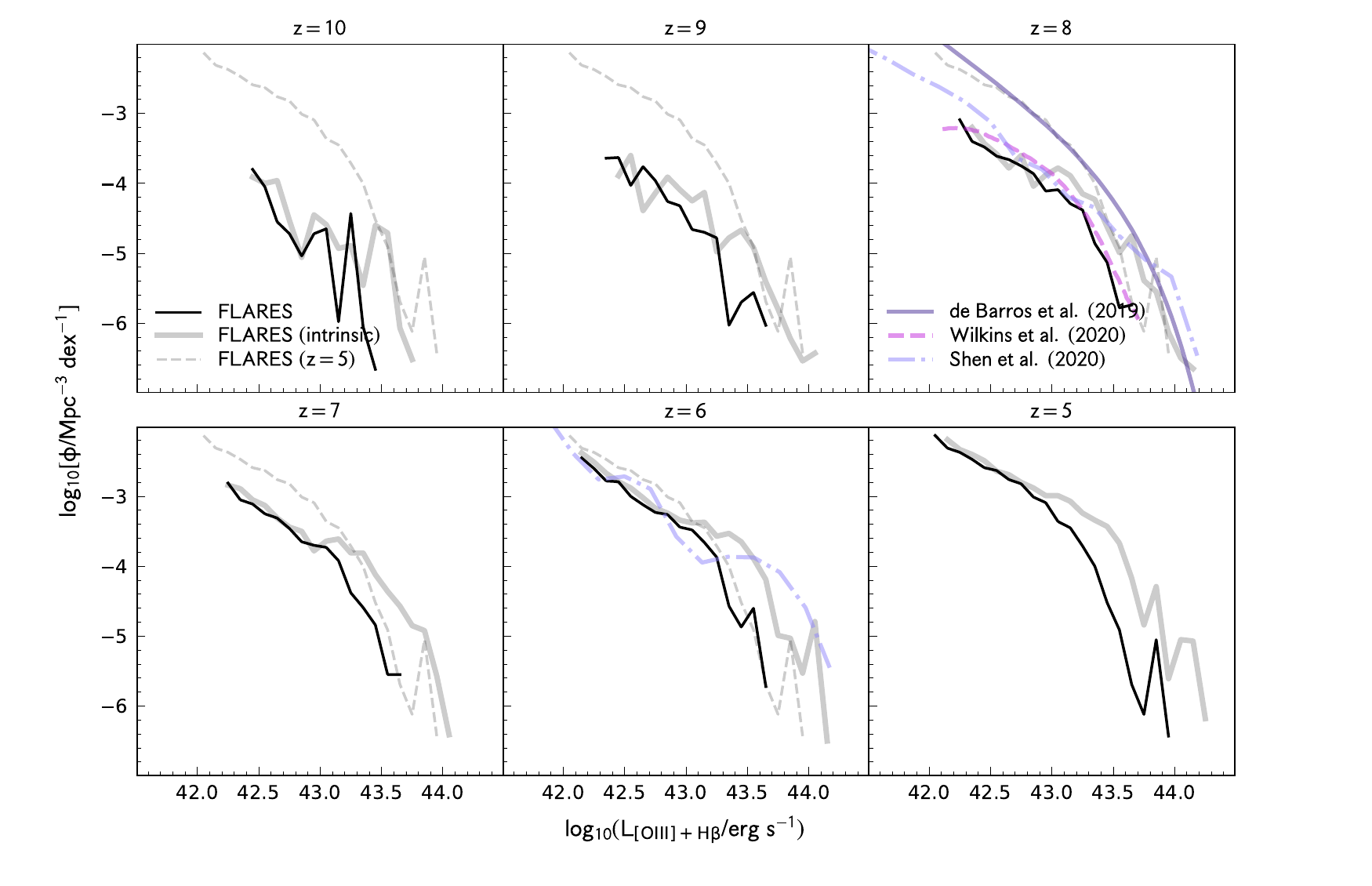}
	\caption{The same as Figure \ref{fig:LF} but instead showing the luminosity function of the combined \oiii\ and H$\beta$ line. Observational constraints from \citet{deBarros19_OIIIHbeta} are shown at $z\approx 8$ alongside model predictions from \citet{Wilkins20} and \citet{Shen2020}.}\label{fig:LF_OIIIHb}
\end{figure*}

We begin by exploring the shape and redshift evolution of the \oiii$\lambda 5008$\AA\ and \oiii\ + H$\beta$ luminosity functions (LF) in Figures \ref{fig:LF} and \ref{fig:LF_OIIIHb} respectively. The shape of LF broadly follows that of UV LF showing a clear drop at high-luminosities, consistent with an exponential like drop-off, and at fainter luminosities a power-law behaviour \citep{FLARES-II}. The UV LF itself tracks the evolution of galaxy stellar mass function \citep[GSMF, see ][]{FLARES-II} but with a steeper drop-off due to the effect of dust attenuation. Figures \ref{fig:LF} and \ref{fig:LF_OIIIHb} also show the intrinsic \oiii\ luminosity function revealing that, like the far-UV luminosity function \citep[see][]{FLARES-II}, the density of the brightest galaxies is suppressed by dust attenuation. The impact of dust attenuation is explored in more detail below in \S\ref{sec:predictions:dust}. Figure \ref{fig:LF} also includes a comparison with observations, this is discussed below in Section \ref{sec:observations}.

\subsection{\oiii--UV luminosity relation}

\begin{figure*}
    \includegraphics[width=2\columnwidth]{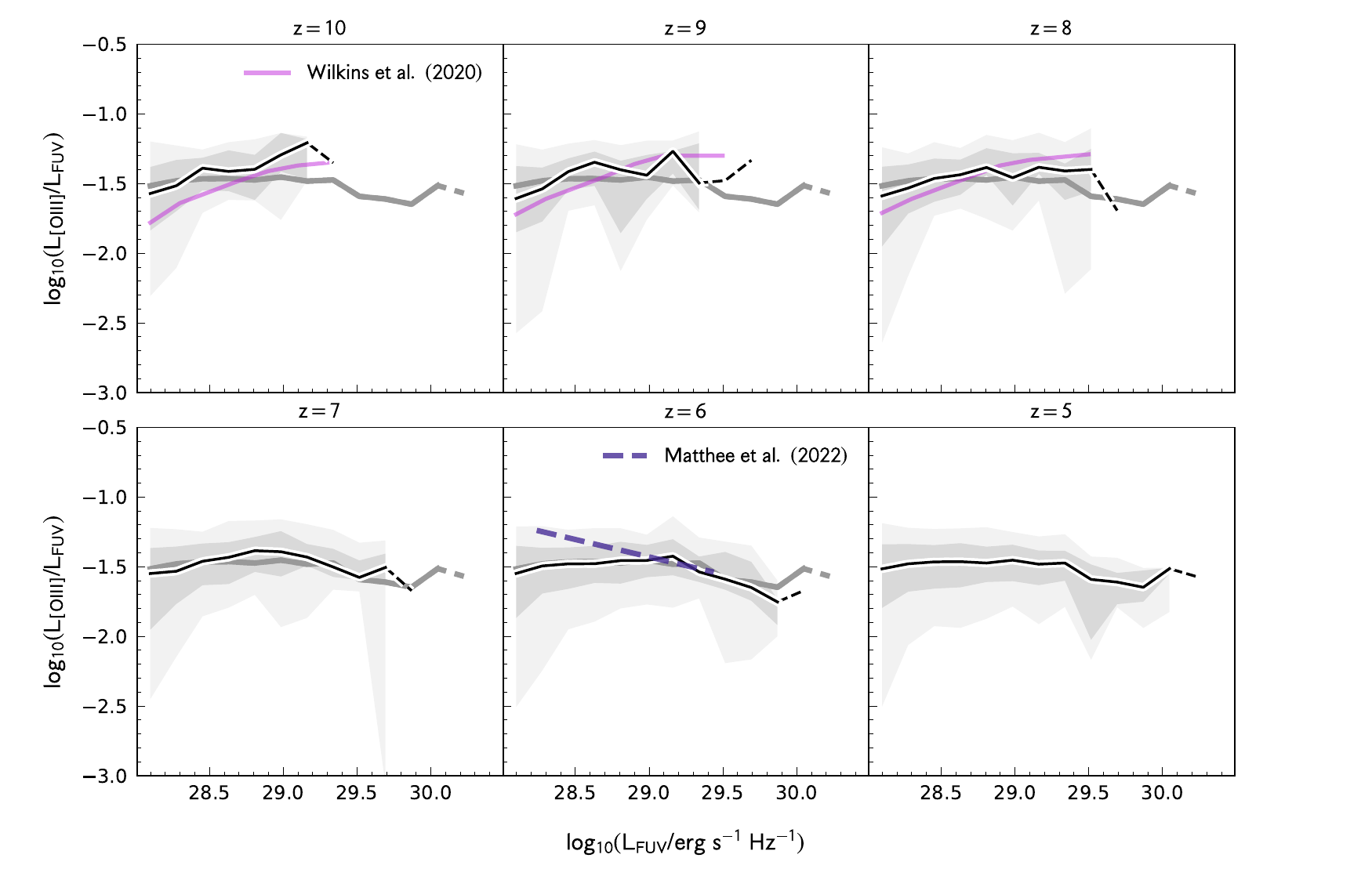}
	\caption{The relationship between the \oiii\ luminosity and the far-UV luminosity, expressed as a ratio, predicted by \flares. The outlined black line show the median \oiii\ luminosity while the grey line shows the median at $z=5$. The two shaded regions show the central 68\% and 95\% ranges. The dashed line shows the relationship obtained by \citet{Matthee23}. The solid blue curve shows predictions from the Bluetides simulation \citep{Wilkins20}. Note: the FUV luminosity included in the line ratio is expressed in units of erg/s not erg/s/Hz. \label{fig:L}}
\end{figure*}

The evolution of the \oiii\ luminosity function across this redshift interval largely tracks that of the rest-frame far-UV luminosity function. To show this more clearly, in Figure \ref{fig:L} we show the relationship between the \oiii\ luminosity and the far-UV luminosity, again from $z=5\to 10$. Firstly, this reveals a flat relationship, i.e. $L_{\oiii}$ tracks $L_{\rm FUV}$. Since the most luminous galaxies in \flares\ have significantly higher metallicity \citep[see][]{FLARES-VII} and that \oiii\ luminosities drop precipitously with metallicity (see Figure \ref{fig:theory_Z}, \S\ref{sec:theory:sfzh}) this is perhaps surprising and we might naively expect to see a drop in $L_{\oiii}/L_{\rm FUV}$ with $L_{\rm FUV}$. However, in \flares\ \oiii\ emission if generally less susceptible to dust than the UV compensating for the metallicity driven drop in \oiii\ emission. Secondly, the relationship between $L_{\oiii}$ and $L_{\rm FUV}$ shows little evolution with redshift. This predominately reflects the fact that the far-UV and LyC emission are both driven by young/massive stars.

\subsection{\oiii\ equivalent width distribution}

\begin{figure*}
    \includegraphics[width=2\columnwidth]{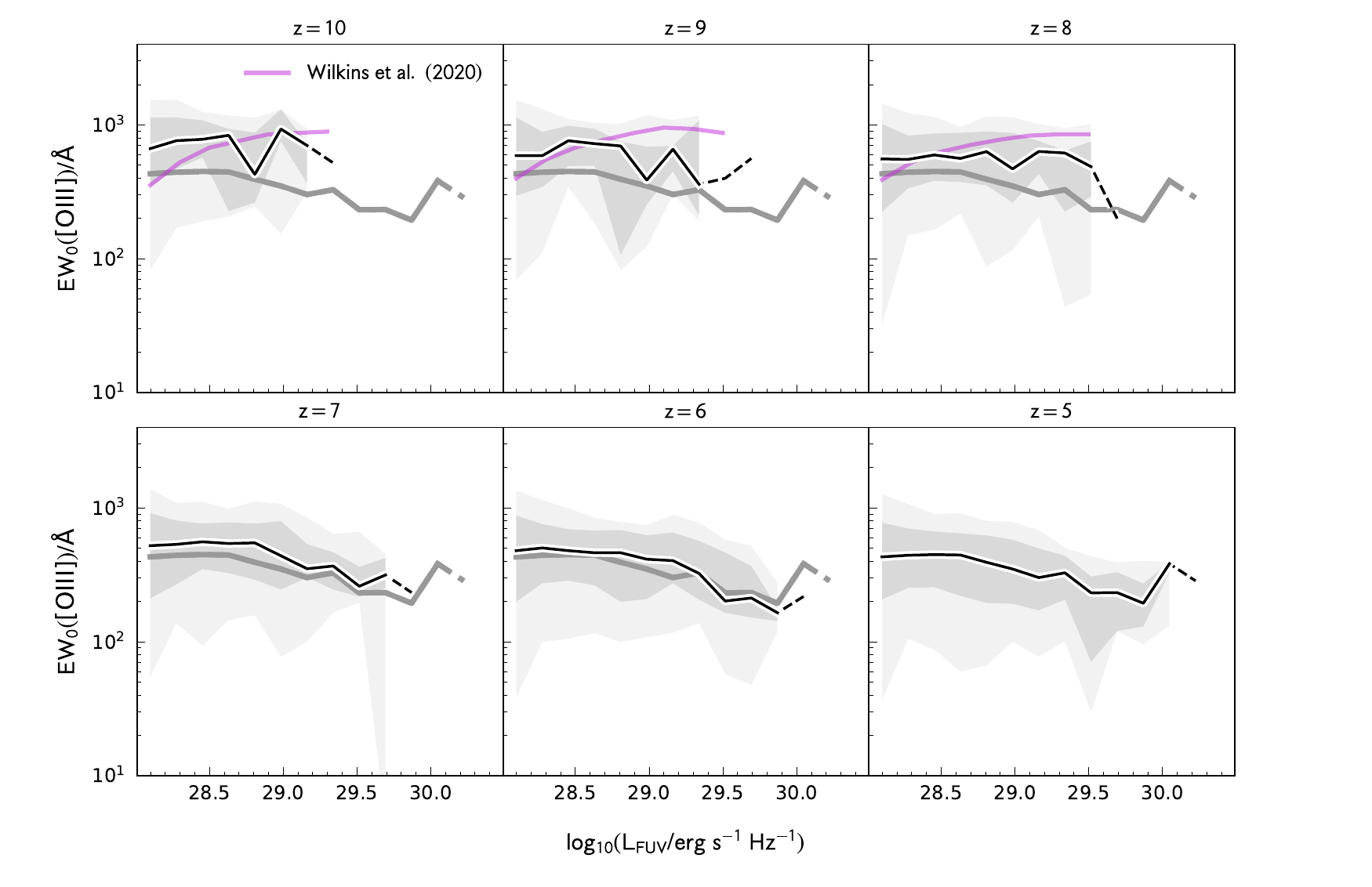}
	\caption{The same as Figure \ref{fig:L} but showing the rest-frame equivalent width of \oiii. \label{fig:EW}}
\end{figure*}

Next, in Figure \ref{fig:EW}, we show predictions for the evolution of the relationship between the rest-frame \oiii\ equivalent width and far-UV luminosity. This reveals a predominantly flat relationship at all redshifts, with a slight decline at the brightest luminosities. This decline is driven by the higher metallicity combined with slightly older ages in these galaxies \citep[see][]{FLARES-VII}. Unlike the $L_{\oiii}$-$L_{\rm FUV}$ relation, which shows little redshift evolution, the average equivalent widths do evolve with redshift, increasing by $\approx 50\%$ from $z=5\to 10$. The reflects the fact that the while the LyC emission is sensitive to only the most recent star formation the optical continuum emission is produced by stars across the preceding few hundred million years.

\subsection{\oiii$\lambda 5008$-\hb\ ratio}

\begin{figure*}
    \includegraphics[width=2\columnwidth]{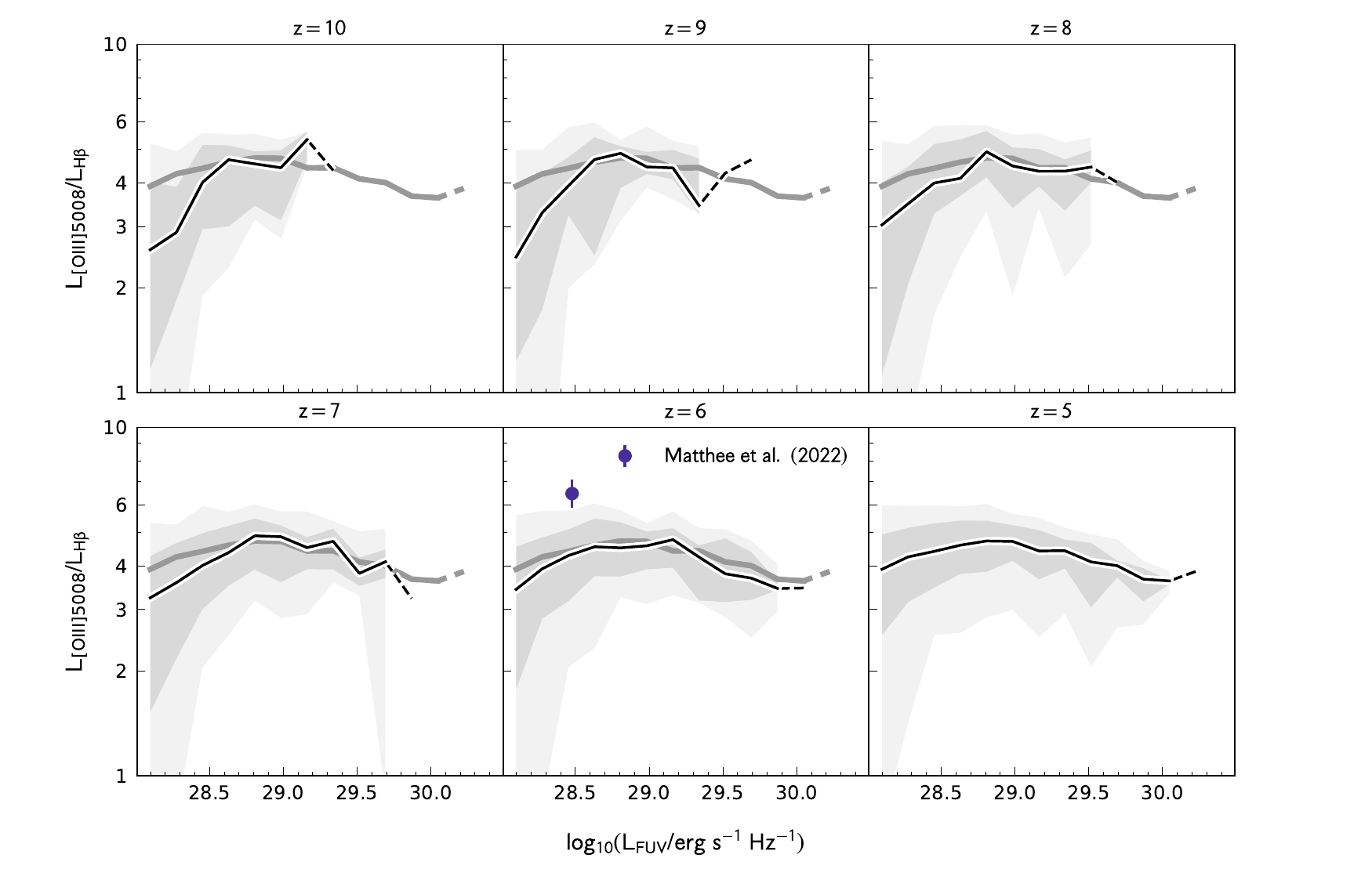}
	\caption{The same as Figure \ref{fig:L} but showing the ratio of the \oiii$\lambda 5008$ to H$\beta$ line luminosities. The point shows the observational constraints of \citet{Matthee23} for their full sample of 117 \oiii\ emitters. \label{fig:lineratio}}
\end{figure*}

An additional useful, observationally accessible, diagnostic is the ratio of the  \oiii$\lambda 5008$ to \hb\ line luminosities. Predictions from \flares\ for this ratio are shown in Figure \ref{fig:lineratio}. Unlike the EW, this ratio is not particularly sensitive to the star formation history, at least for actively star forming galaxies. The ratio is however sensitive to extreme metallicities ($Z<0.001$, $Z>0.01$,) where it rapidly drops, as well as the ionisation parameter $U$. As the majority of \flares\ galaxies span the range $Z=0.001-0.01$ and we assume a single reference ionisation parameter (and thus have a narrow range of effective ionisation parameters) it is not surprising that the \flares\ predictions are tightly clustered around $\approx 5$.

\subsection{Impact of reprocessing by dust}\label{sec:predictions:dust}

\begin{figure}
    \includegraphics[width=\columnwidth]{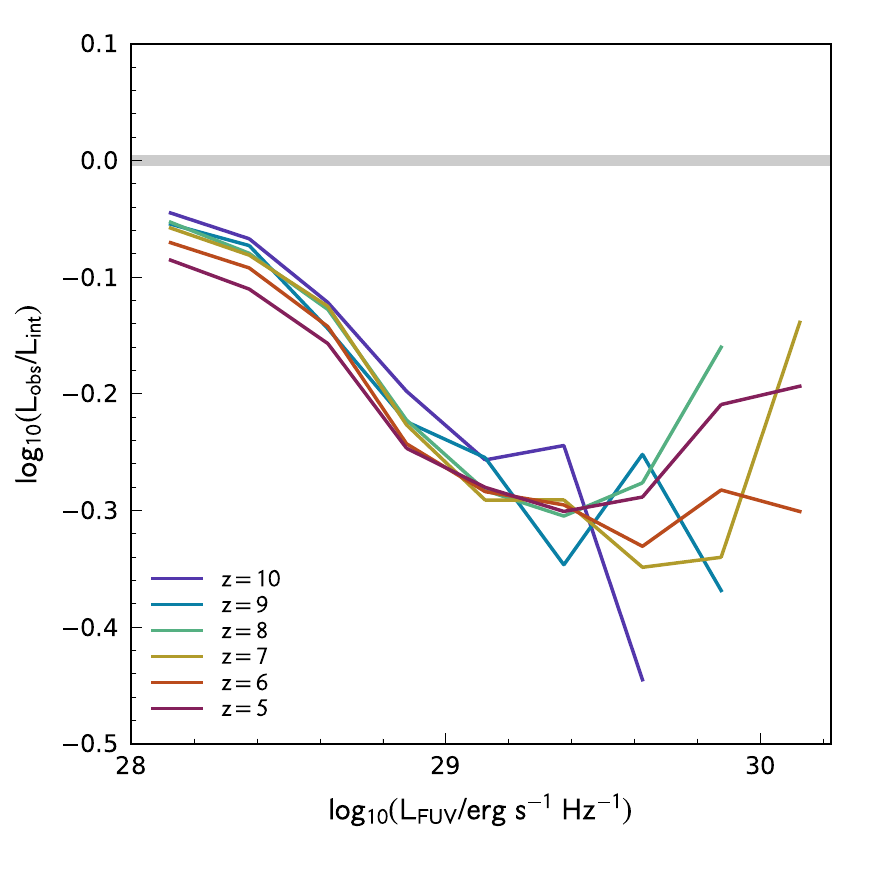}
    \includegraphics[width=\columnwidth]{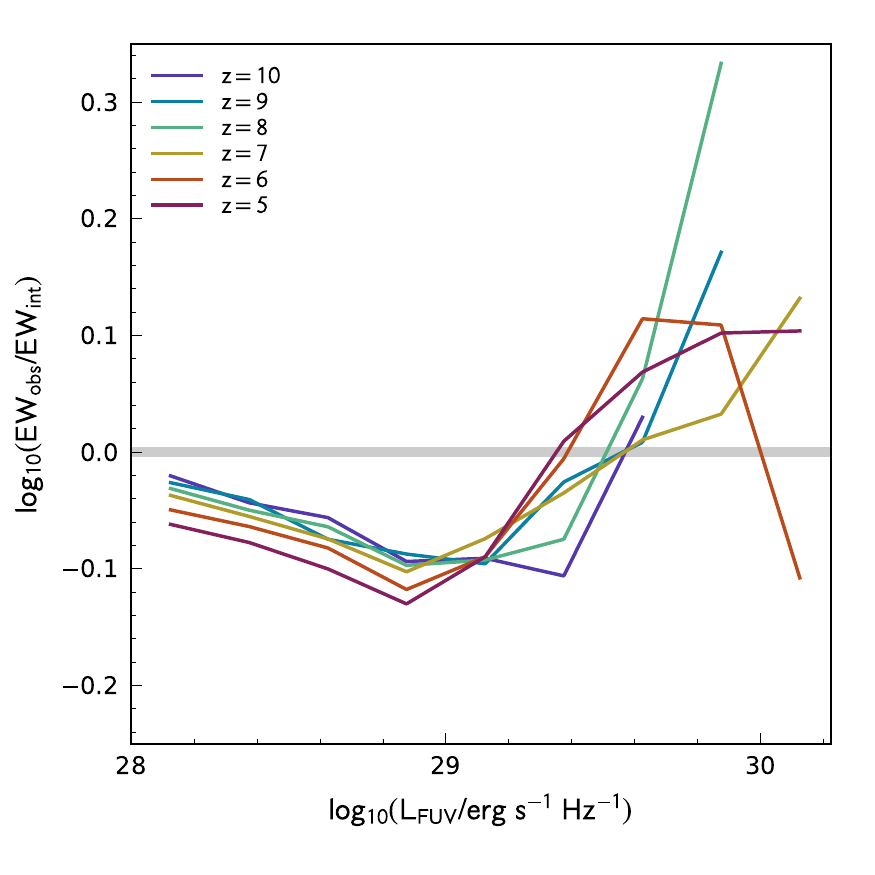}
	\caption{The average (median) impact of dust attenuation of the \oiii\ luminosity (top) and EW (bottom), both expressed as a function of the observed far-UV luminosity.  \label{fig:reprocessing}}
\end{figure}

To further explore the impact of dust on our predictions, in Figure \ref{fig:reprocessing} we show the ratio of the intrinsic to attenuated line luminosities (top panel) and equivalent widths (bottom) as a function of the dust-attenuated far-UV luminosity.

As already hinted at in our comparison of the attenuated and intrinsic luminosity functions dust attenuation increases with both the far-UV and \oiii\ luminosity. Compared to the intrinsic luminosity (not shown) the attenuation continues to increase with increasing luminosity. However, compared to the dust-attenuated (observed) luminosity the attenuation flattens since the most heavily obscured galaxies have lower observed luminosities. In the context of \flares\ dust attenuation is due to both a birth cloud component, linked to the metallicity of each star particle, and an ISM component linked to the intervening surface density of metals. The more massive (and intrinsically bright) a galaxy generally the higher the stellar \citep[see][]{FLARES-VIII} and gas-phase metallicity. More massive galaxies also have larger gas reservoirs and thus metal (and dust) surface densities.

The effect of dust attenuation on equivalent widths is more complex. Our faintest galaxies show mild suppression peaking at $L_{\rm FUV}\approx 10^{29}\ {\rm erg\ s^{-1}\ Hz^{-1}}$ where the EW is reduced by 0.1 dex. At brighter far-UV luminosities the impact of dust on the EW dust declines and eventually leads to a small enhancement of EWs in the most UV luminous galaxies. The interpretation here is that in fainter galaxies, young stellar populations, which dominate the production of LyC photons and thus the \oiii\ line luminosity, are affected by slightly higher dust attenuation than the populations that give rise to the continuum emission. In the context of the \flares\ model this is expected due to the addition of a birth cloud dust component. In the most luminous galaxies, which also roughly corresponds to the most attenuated systems, this additional birth cloud is sub-dominant to dust in the wider ISM. In these galaxies, the enhancement of EWs relative to the intrinsic values is   explained by the fact that the \oiii\ producing stellar populations are preferentially found on the outskirts of galaxies compared to the continuum generating populations.

\subsection{Correlation with Physical Properties}\label{sec:predictions:physical}

\begin{figure}
    \includegraphics[width=\columnwidth]{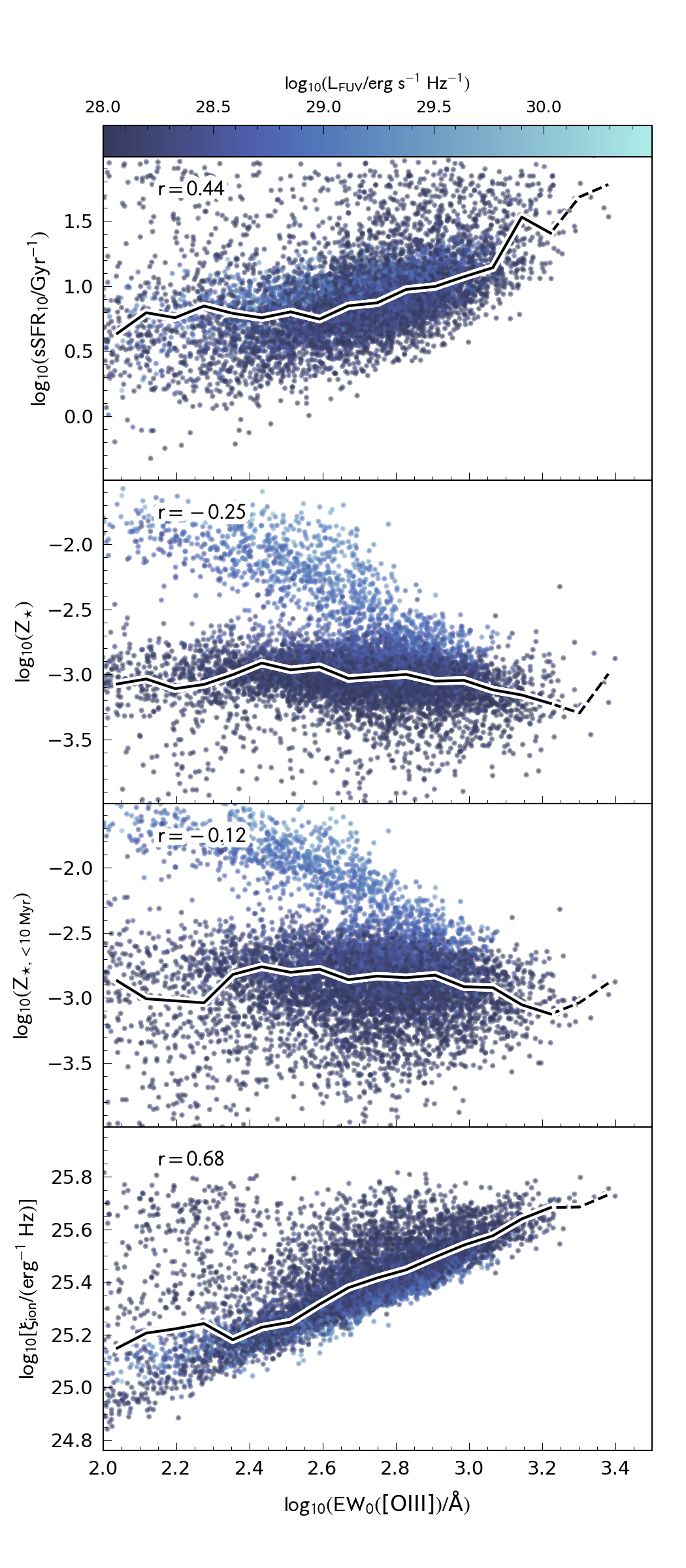}
	\caption{Comparison between the rest-frame \oiii\ EW and the specific star formation rate, stellar metallicity (both total and young), and the ionising photon production efficiency $\xi_{\rm ion}$. The star formation rate is averaged over the preceding 10 Myr while the stellar metallicity is defined for all stellar populations. \label{fig:physical}}
\end{figure}

We next explore, in Figure \ref{fig:physical}, how the \oiii\ equivalent width correlates with key physical properties, including the specific star formation rate, total stellar metallicity, the stellar metallicity of young stellar populations, and the ionising photon production efficiency $\xi_{\rm ion}$. Here, the star formation rate is defined as the mass of stars that have formed in the last 10 Myr, the stellar metallicity is the mass weighted stellar metallicity, and, in common with most observational studies, $\xi_{\rm ion}$ as the ratio of the Lyman continuum (ionising) photon production rate ($\dot{n}_{\rm LyC}$) to the rest-frame observed UV luminosity.

This reveals a correlation, albeit relatively weak ($r=0.44$), with specific star formation rate, such that galaxies with the highest \oiii\ EW generally have higher specific star formation rates. This is expected since \oiii\ emission is driven by young stars while the optical continuum includes a contribution from older stellar populations. 

The relationship between the \oiii\ EW and stellar metallicity is more complex, with two clear branches. Because of the steep shape of the galaxy stellar mass function, and the existence of a tight mass--metallicity relation in \flares, the vast majority of our galaxies have $Z_{\star}=0.001$. These faint, low-mass galaxies exhibit a range of EWs spanning $\approx 100-2000\ {\rm\AA}$ with the scatter driven by several effects, including the star formation history, dust, but crucially the metallicity of the \oiii\ population, not just the overall metallicity. The second, upper, branch corresponds to more massive, luminous galaxies with higher-metallicities. The metallicities of the \oiii\ producing stellar populations in these galaxies falls beyond the peak in the \oiii\ luminosity--metallicity relation (see Figure \ref{fig:theory_Z}). Since this branch corresponds to the most massive and luminous systems it is also strongly impacted by dust attenuation, which has the effect of increasing the scatter in the EW but, as found previously, not significantly reducing it.

Finally, in the bottom panel of Figure \ref{fig:physical}, we show the relationship between the EW and the ionising photon production efficiency. This shows a clear correlation ($r=0.68$) such that the most extreme emitters have the largest production efficiencies. This is of course not a surprise considering \oiii\ line emission is, at least in \flares, driven by the production of ionising photons by massive stars.

\section{Comparison with Observational Constraints}\label{sec:observations}

We now turn our attention to a comparison with recent observational constraints from \hubble, \spitzer, and \jwst, including \citet{deBarros19_OIIIHbeta},  \citet{Endsley22}, \citet{Matthee23}, and \citet{Sun23}. The former two studies infer the \oiii\ + \hb\ properties from broadband photometry, while the latter two use spectroscopic observations recently obtained by \jwst.

\subsection{Photometric Constraints}

The predicted strength of the combination of the \oiii\ and \hb\ emission is sufficient to significantly boost broadband fluxes by up-to $\approx 0.3$ dex \citep[e.g.][]{Wilkins2013d, Wilkins20}. With one band encompassing the line emission and a second probing the (strong line emission free) continuum, it is then possible to infer the luminosities and equivalent widths of these combined lines. In particular, combining \hubble\ and \spitzer/IRAC observations \citep[e.g.][]{Smit2014, Roberts_Borsani_2016, deBarros19_OIIIHbeta, Endsley2021} or, more recently, \jwst\  \citep[e.g.][]{Endsley22}, it is possible to constrain the combined \oiii\ and \hb\ emission.

\begin{figure}
    \includegraphics[width=\columnwidth]{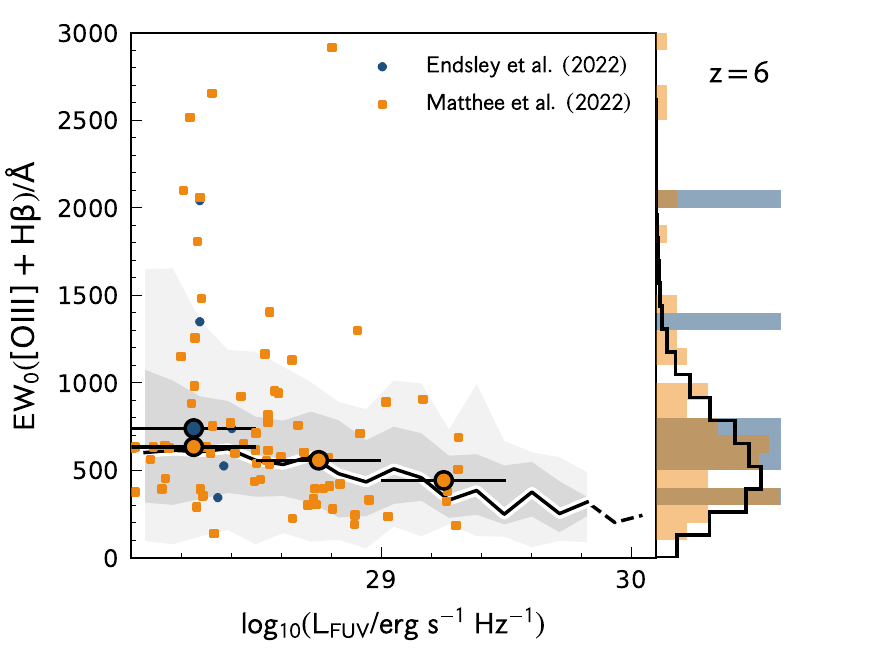}
    \includegraphics[width=\columnwidth]{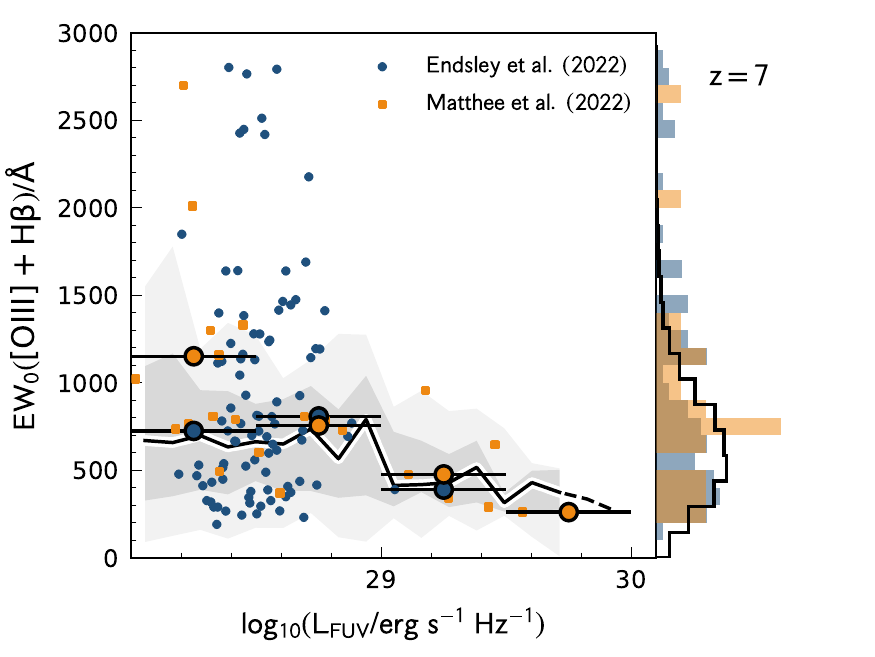}
    \includegraphics[width=\columnwidth]{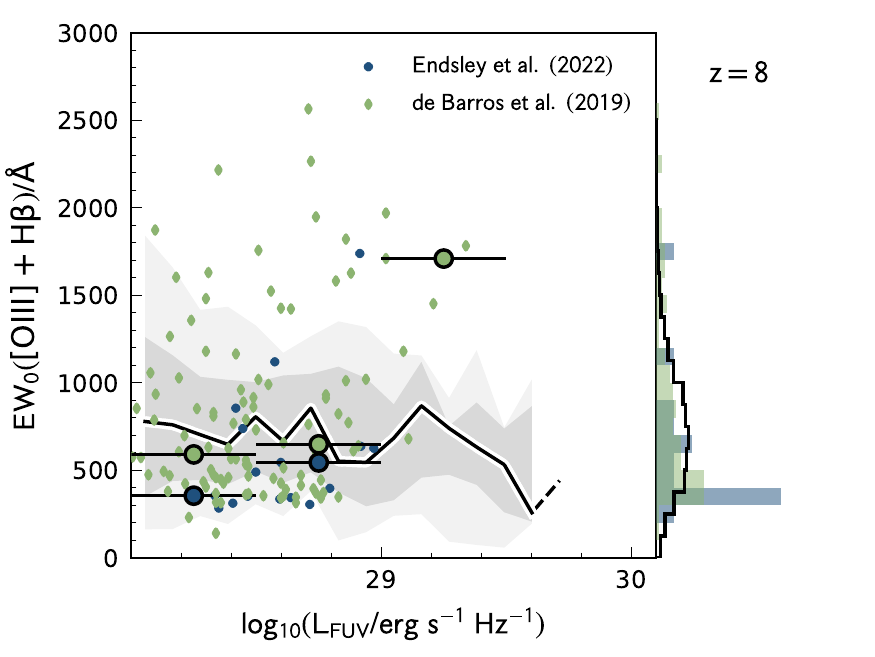}
	\caption{Comparison between our predicted \oiii\ + \hb\ equivalent widths and observations from \citet{deBarros19_OIIIHbeta}, \citet{Endsley22}, and \citet{Matthee23} at $z=5-9$. \label{fig:observations}}
\end{figure}

In Figure \ref{fig:observations} we compare the combined \oiii\ + \hb\ EWs predicted by \flares\ with photometric constraints from \citet{deBarros19_OIIIHbeta} and \citet{Endsley22} at $z=6-9$. In both cases we find good agreement between the predicted and observed median EWs. This supports the findings of \citet{FLARES-VI} where we made a direct comparison between the broadband \hubble\ and \spitzer\ colours of observed galaxies and galaxies predicted by \flares, finding good agreement. However, while we broadly reproduce the median EW, we fail to reproduce the high-EW tail observed in both \citet{deBarros19_OIIIHbeta} and \citet{Endsley22}. In the context of \oiii\ driven by stellar populations there is little remaining model flexibility, since our assumptions tend to maximise the possible EWs. For example, assuming an alternative reference ionisation parameter $U$ would not significantly increase EWs since our adopted value ($0.01$) maximises the EW, as demonstrated in Figure \ref{fig:theory_U} (\S\ref{sec:theory:geometry}). This suggests the explanation lies either with the observations themselves, or there is a significant additional source of ionising photons present in high-redshift galaxies. While \flares\ predicts AGN are present in these galaxies, their contribution to the LyC luminosity is, on average, relatively small (see Kuusisto et al., \emph{in-prep}). \citet{deBarros19_OIIIHbeta} also present a measurement of the combined \oiii\ and H$\beta$ luminosity function, this is shown, alongside \flares\ predictions in Figure \ref{fig:LF_OIIIHb}. \citet{deBarros19_OIIIHbeta} generated this by combining the observed far-UV luminosity function with the measured combined line luminosity - far-UV luminosity relation. This measurement is significantly higher than our predictions, which is surprising considering the otherwise good agreement with typical equivalent widths. A similar disagreement was noted by other theoretical models \citep{Wilkins20, Shen2020}, see \S\ref{sec:models}. One potential explanation here is the accounting of dust.

\subsection{Spectroscopic constraints}\label{sec:observations.spectroscopic}

As noted in the introduction, a handful of spectroscopic constraints on \oiii\ are now available at $z>6$. At present these samples are small, but will rapidly grow thanks to large spectroscopic programmes such as JADES and CEERS. 

\citet{Sun23} presented observations of four serendipitously discovered emission line galaxies at  $z = 6.11 - 6.35$ in \jwst\ NIRCam wide field slitless spectroscopy (WFSS)  comissioning data. These galaxies exhibit \oiii 5008\AA\ EWs ranging from $100-1200{\rm\AA}$ with a median \oiii\ + \hb\ EW $546\pm 77$\AA, consistent with our predictions of $\approx 500-600{\rm\AA}$ at the same redshift and luminosity range. \citet{Sun23} also present constraints on the \oiii 5008\AA\ luminosity function, and these are presented in Figure \ref{fig:LF}. These are somewhat higher than the predictions from \flares, but the sample size is small and consequently the statistical uncertainties are very large. Since the observations are based on a single NIRCam pointing there is also the possibility of significant field-to-field variation \citep[e.g][]{FLARES-X}.

More recently, \citet{Matthee23} presented spectroscopic constraints from the \jwst\ NIRCam WFSS Emission-line galaxies and Intergalactic Gas in the Epoch of Reionization \citep[EIGER,][]{Kashino22} survey. \citet{Matthee23} present both pure spectroscopic constraints alongside spectro-photometric constraints, combining line fluxes from the WFSS with photometry from the NIRCam imaging. 

The combined \oiii\ + \hb\ equivalent widths of the \citet{Matthee23} observations are shown in Figure \ref{fig:observations}. Here we choose to compare with the spectro-photometric constraints since these allow a better estimation of the continuum flux than the pure spectroscopic observations alone. We find good agreement between the predicted and observed median of the EW distribution. However, like with the photometric constraints we find an excess in the number of high-EW sources. 

\citet{Matthee23} also constrain the relationship between the far-UV and \oiii\ luminosity, shown in Figure \ref{fig:L} and the \oiii 5008\AA\ luminosity function, shown in Figure \ref{fig:LF}. The normalisation of the $L_{\rm FUV}$-$L_{\oiii}$ relationship matches that predicted by \flares, though the slope is somewhat steeper, resulting in fainter galaxies having higher ratios than predicted by \flares. At faint ($L_{\oiii}<10^{43}\ {\rm erg/s}$) luminosities the \flares\ predictions provide an excellent match to EIGER luminosity function. However, at brighter luminosities the EIGER constraints tend to be higher than predicted by \flares, suggesting an excess of bright \oiii\ emitting galaxies. One possible explanation here is cosmic variance. Indeed, the EIGER field contains an over-density at $z\approx 6.77$ to which three of four sources with $M_{\rm FUV}<-22$ belong. While the EIGER constraints lie above our predictions, which include dust attenuation, they are however comparable to our intrinsic predictions. Another possibility is then that we have over-predicted the amount of dust attenuation in these systems.

Spectroscopic observations also have the advantage that they can separate the contributions of \oiii\ and \hb, allowing us to measure line ratios.  {\citet{Matthee23} measure \oiii 5008/\hb\ (R3) line ration in their sample, finding an average value of $6.3$. This is slightly larger than our predicted typical values ($4-5$) possibly suggesting the need for higher ionisation parameters (see \S\ref{sec:theory:geometry}).

\section{Comparison with other models}\label{sec:models}

Emission line predictions of high-redshift galaxies have previously been made for other cosmological hydrodynamical simulations, including \textsc{MassiveBlack} \citep{Wilkins2013d}, \textsc{Bluetides} \citep{Wilkins20}, \textsc{Illustris-TNG} \citep{Shen2020, Hirschmann2023}, and \textsc{Simba} \citep{Garg2022}. Here we compare our predictions with these studies where they present overlapping quantities and make available their data.

\citet{Wilkins20} models a selection of UV and optical lines for galaxies in the Bluetides simulation \citep{Bluetides-I, Bluetides-II} at $z=8-13$. Bluetides is a single very large [$(400/h\ {\rm Mpc})^3$] periodic volume with a resolution, quantified by the gas-particle mass, slightly below that of \eagle/\flares, and utilises a distinct physics model. The Bluetides nebular emission and dust modelling was very similar to that used in this work. The predicted relationships between the far-UV luminosity and the $\oiii$ luminosity and equivalent width at $z=8-10$ are shown in Figures \ref{fig:L} and \ref{fig:EW} respectively. In both cases the Bluetides predictions are similar to those from \flares. However, in both cases Bluetides predicts a clear increase with $L_{\rm FUV}$, at least at $L<10^{29}\ {\rm erg\ s^{-1}\ Hz^{-1}}$. Similarly, the $\oiii\lambda$5008\AA\ luminosity function predicted by Bluetides shows excellent agreement with our predictions (see Figure \ref{fig:LF}).

\citet{Shen2020} model the H$\alpha$, H$\beta$, and \oiii\ lines, amongst other quantities, of galaxies at $z=2-8$ using the Illustris-TNG \citep{IllustrisTNG} simulation suite\footnote{The suite includes three simulations with varying volume and resolution. The middle $(100\ {\rm Mpc})^3$ is comparable in resolution to \flares, while the larger (smaller) volumes are lower (higher) resolution.}. \citet{Shen2020} modelled lines using \textsc{Mappings iii} \citep{Groves2008} at $z<6$ and \textsc{fsps} \citep{FSPSI, Byler17} above. \citet{Shen2020} present  predictions for the \oiii\ + H$\beta$ luminosity function at $z=6$ and $z=8$. These are shown, alongside our predictions, in Figure \ref{fig:LF_OIIIHb}.  At $L=10^{42-43}\ {\rm erg\ s^{-1}}$, where the bulk of our objects lie, the agreement is good. At brighter luminosities however the source density predicted by \citet{Shen2020} is significantly larger than predicted by \flares. While \citet{Shen2020} do not present predictions for \oiii$\lambda 5008$\AA\ alone it is likely that it follows a similar pattern suggesting a better agreement with the recent \citet{Matthee23} observational constraints.


\section{Conclusions}\label{sec:conc}

In this work we have explored the \oiii\ properties of the galaxy population in the First Light And Reionisation Epoch Simulations (\flares). The \flares\ strategy enables us to predict the properties of galaxies over a wide range of masses and luminosities at high-redshift.

Our main conclusions are:

\begin{itemize}
    \item The \oiii\ luminosity function (LF) predicted by \flares\ declines sharply with redshift with the density of sources dropping by $\approx 1$ dex from $z=5$ to $10$.  The bright-end of the LF is strongly impacted by dust with a suppression of 1 dex at $L\sim 10^{43.5}\ {\rm erg/s}$. While the faint end of the LF is well matched to recent spectroscopic constraints \citep[][]{Sun23, Matthee23} we predict fewer very bright sources than \citet{Matthee23} with one possible explanation being comsic variance.
    
    \item We predict a flat, un-evolving, relationship between the far-UV and \oiii\ line luminosities. While the intrinsic ratio falls to higher luminosity due to the effect of increasing metallicity this is moderated by the growing impact of dust.
    
    \item We predict a median \oiii\ rest-frame equivalent width (EW) of $\approx 500$\AA\ at $z=5$. This declines slightly with far-UV luminosity and stellar mass and increases to higher-redshift. At low UV luminosities dust-attenuated EWs are slightly smaller than the intrinsic values. However, at higher luminosities ($M_{\rm FUV}<-21$), dust-attenuated EWs are slightly higher than the intrinsic values indicating that the continuum emission is more heavily attenuated than the line. The interpretation here is that the young \oiii\ producing stellar population are preferentially found on the outskirts of galaxies compared to the continuum producing population. We find that the \oiii\ EW correlates weakly with specific star formation rate but more strongly with ionising photon production efficiency. The relationship with metallicity is more complex with two clear branches.
    
    \item Our median EWs are consistent with both recent photometric \citep[][]{deBarros19_OIIIHbeta, Endsley22} and spectroscopic \citep{Sun23, Matthee23} constraints. However, we fail to predict the tail of galaxies with extremely high ($>2000$\AA) EWs found by these studies possibly suggesting an additional source of ionising photons in these systems.
    
    \item We predict \oiii 5008\AA/H$\beta$ ratios of $\approx 4-5$, slightly smaller than those found by \citet{Matthee23}. However, our ratios are strongly affected by our assumed reference ionisation parameter.
    
\end{itemize}

Spectroscopic constraints of galaxies in the distant Universe will imminently be transformed by surveys such as JADES and CEERS. Together with other cycle 1/2 observations these will vastly increase the sample size and dynamic range of spectroscopic observations of \oiii\ and other lines yielding new constraints on physical models in this epoch of the Universe's history.

\section*{Acknowledgements}

We thank the \eagle\, team for their efforts in developing the \eagle\, simulation code. We wish to thank Scott Kay and Adrian Jenkins for their invaluable help getting up and running with the \eagle\, resimulation code. 

This work used the DiRAC@Durham facility managed by the Institute for Computational Cosmology on behalf of the STFC DiRAC HPC Facility (www.dirac.ac.uk). The equipment was funded by BEIS capital funding via STFC capital grants ST/K00042X/1, ST/P002293/1, ST/R002371/1 and ST/S002502/1, Durham University and STFC operations grant ST/R000832/1. DiRAC is part of the National e-Infrastructure. We also wish to acknowledge the following open source software packages used in the analysis: \textsc{Scipy} \cite[][]{2020SciPy-NMeth}, \textsc{Astropy} \cite[][]{robitaille_astropy:_2013}, \textsc{Matplotlib} \cite[][]{Hunter:2007} and WebPlotDigitizer \cite[][]{Rohatgi2020}. 

APV acknowledges support from the Carlsberg Foundation (grant no CF20-0534). PAT acknowledges support from the Science and Technology Facilities Council (grant number ST/P000525/1). DI acknowledges support by the European Research Council via ERC Consolidator Grant KETJU (no. 818930). CCL acknowledges support from a Dennis Sciama fellowship funded by the University of Portsmouth for the Institute of Cosmology and Gravitation. The Cosmic Dawn Center (DAWN) is funded by the Danish National Research Foundation under grant No. 140.

We list here the roles and contributions of the authors according to the Contributor Roles Taxonomy (CRediT)\footnote{\url{https://credit.niso.org/}}.
\textbf{Stephen M. Wilkins}: Conceptualization, Data curation, Methodology, Investigation, Formal Analysis, Visualization, Writing - original draft.
\textbf{Christopher C. Lovell,  Aswin P. Vijayan}: Data curation, Methodology, Writing - review \& editing.
\textbf{Nathan Adams, Joseph Caruana, Chris Concelice, James Donnellan, Dimitrios Irodotou, Jorryt Matthee, Pablo G. P\'erez-Gonz\'alez, William Roper, Louise Seeyave, Jack Turner, Aprajita Verma}: Writing - review \& editing.

\section*{Data Availability}

The data associated with the paper will be made publicly available at \href{https://flaresimulations.github.io/data.html}{https://flaresimulations.github.io/data.html} on the acceptance of the manuscript.



\bibliographystyle{mnras}
\bibliography{flares, flares-oiii,flares-oiii-jwst} 





\bsp	
\label{lastpage}
\end{document}